\documentclass[runningheads]{llncs}
\usepackage[T1]{fontenc}
\usepackage{graphicx}
\usepackage[letterpaper]{geometry}
\usepackage{paralist}
\usepackage{hyperref}

\usepackage{amsmath,amssymb,amsfonts}
\usepackage{xspace}
\usepackage[inline]{enumitem}
\usepackage{graphicx}
\usepackage{tabularx}
\usepackage[multiple]{footmisc}
\usepackage{subcaption}
\usepackage{tikz}
\usepackage{tabularx,booktabs}
\usepackage{diagbox}
\usepackage{colortbl}
\usepackage{float}
\usepackage{footnote}

\usepackage{diagbox}
\usepackage{bbding}
\usepackage{dsfont}
\usepackage[ruled,vlined,linesnumbered]{algorithm2e}
\usepackage{makecell}
\usepackage{multirow}
\usepackage{multicol}
\usepackage{xcolor}

\usepackage{wrapfig}

\newcommand{\fgr}[3][\relax]{%
	\begin{figure}[htp]%
		\centering
		\includegraphics[#2]{#3}%
		\ifx\relax#1\else\caption{{#1}}\fi
	\end{figure}%
}

\newcommand{\hide}[1]{}

\newcommand{\method}{{\sc LM-Gsum}\xspace}

\newcommand{\etal}{{\it et al.}}

\newcommand{\redlinedashed}{\raisebox{2pt}{\tikz{\draw[-,red,dashed,line width = 1pt](0,0) -- (4mm,0);}}}

\newcommand{\redlinesolid}{\raisebox{2pt}{\tikz{\draw[-,red,line width = 1pt](0,0) -- (4mm,0);}}}

\newcommand{\cbit}{\begin{compactitem}}
	\newcommand{\ceit}{\end{compactitem}}
\newcommand{\cben}{\begin{compactenum}}
	\newcommand{\ceen}{\end{compactenum}}

\makeatletter
\newcommand\footnoteref[1]{\protected@xdef\@thefnmark{\ref{#1}}\@footnotemark}
\makeatother

\begin{document}

\hyphenation{Definition}
\hyphenation{Definitions}

\title{Summarizing Labeled Multi-Graphs
}

\author{Dimitris Berberidis\inst{1} \quad Pierre J. Liang\inst{2} \quad Leman Akoglu\inst{1} 
}

\institute{Carnegie Mellon University, 
 Heinz College of Information Systems and Public Policy \\
\email{\{dbermper, lakoglu\}@andrew.cmu.edu}
\and 
Carnegie Mellon University, 
 Tepper School of Business \\ \email{liangj@tepper.cmu.edu}}

%
% \authorrunning{G. D. F. Silva et al.}
%\institute{}
%\email{guilherme.domingos.silva@usp.br, robson@icmc.usp.br} \and
%Carnegie Mellon University, USA\\
%\email{lakoglu@andrew.cmu.edu}}
%
\maketitle              % typeset the header of the contribution
\begin{abstract}
Real-world graphs can be difficult to interpret and visualize beyond a certain size. To address this issue,  {graph summarization} aims to  simplify and shrink a graph, while maintaining its high-level structure and characteristics.
Most summarization methods are designed for homogeneous, undirected, simple graphs; however, many real-world graphs are
\emph{ornate}; with characteristics including node labels, directed edges, edge multiplicities, and self-loops. 
In this paper we propose \method, a \emph{versatile} yet rigorous graph summarization model that (to the best of our knowledge, for the first time) can handle graphs with \textit{all} the aforementioned characteristics (and \textit{any} combination thereof). Moreover, our proposed model captures basic sub-structures that are prevalent in real-world graphs, such as cliques, stars, etc. \method  compactly quantifies the information content of a complex graph using a novel encoding scheme, where %and relying on the Minimum Description Length principle, 
it seeks to minimize the total number of bits required to encode ($i$) the summary graph, as well as  ($ii$) the corrections required for reconstructing the input graph {\em losslessly}.   
To accelerate the summary construction, it creates super-nodes efficiently by merging nodes \textit{in groups}. %, which are efficiently identified using incremental and filtered Locality-Sensitive Hashing.
%Our qualitative 
Experiments demonstrate that \method facilitates the visualization of real-world complex graphs,  revealing interpretable structures and high-level relationships. Furthermore, %quantitative results show that 
\method achieves better trade-off between compression rate and running time, relative to existing methods (only) on comparable settings.   
\end{abstract}

\section{Introduction} \label{01intro}

\label{01intro}
Given a directed labeled multi-graph $G$,
how can we construct a small summary graph $g$ that reflects the high-level structures and relationships in $G$?
How can we find a succinct $g$ that is yet an accurate representation, which requires a small amount of corrections to recover the original $G$?
%What and why graph summarization
With the advent  of technology, not only the size but also the complexity of real-world graphs have grown immensely.
Today graph data often contains node labels, multi-edges, etc. 
Graph summarization aims to find high-level structural patterns and most salient information in large complex graphs to enable efficient storage, processing, visualization and interpretation.

A large body of existing graph summarization techniques is for \textit{plain} graphs with homogeneous
unlabeled nodes {\cite{khan2015set,shin2019sweg,grass,vog,quality_summary,weighted_compression,navlakha}}.
However there exist numerous real-world graphs with multiple node labels;
including transaction networks containing nodes (i.e. accounts) of
various types (cash, revenue, expense, etc.) or
heterogeneous graphs such as publication records among entities of various
types (paper, author, venue, etc.).
We refer to both kinds as node-labeled, or simply \textit{labeled graphs}.
Moreover, a vast majority of prior work are for summarizing \textit{simple} {\cite{khan2015set,shin2019sweg,grass,vog,navlakha,zhu2016unsupervised,liu2012approximate,k_snap,subdue}}
\textit{undirected} {\cite{khan2015set,shin2019sweg,grass,vog,quality_summary,weighted_compression,navlakha,zhu2016unsupervised,k_snap}} graphs, whereas the edges in real-world graphs  may repeat (e.g., multiple transactions between two accounts, multiple exchanges between two email addresses, etc.) which are called \textit{multi-graphs}. As is the case for transaction and email graphs, among others, the edges can also be \textit{directed}.

In this work we propose (to the best of our knowledge; see, Table \ref{tab:salesman}) the \textit{first} method called \method, for {\sc L}abeled {\sc M}ulti-{\sc G}raph {\sc Sum}marization, with directed edges and possible self-loops. (See Sec. \ref{04related}, and a recent comprehensive survey \cite{journals/csur/LiuSDK18}.) 
Besides, \method is \textit{versatile} in that it can also handle graphs with any combination of those properties (i.e., (un)directed, plain/labeled, simple/multi- or weighted graphs).

\begin{wrapfigure}{l}{0.65\textwidth}
%\begin{wrapfigure}[!t]
	%\begin{tabular}{ccccc}
	\vspace{-0.4in}
	\centering
	\includegraphics[width=100mm,height=60mm]{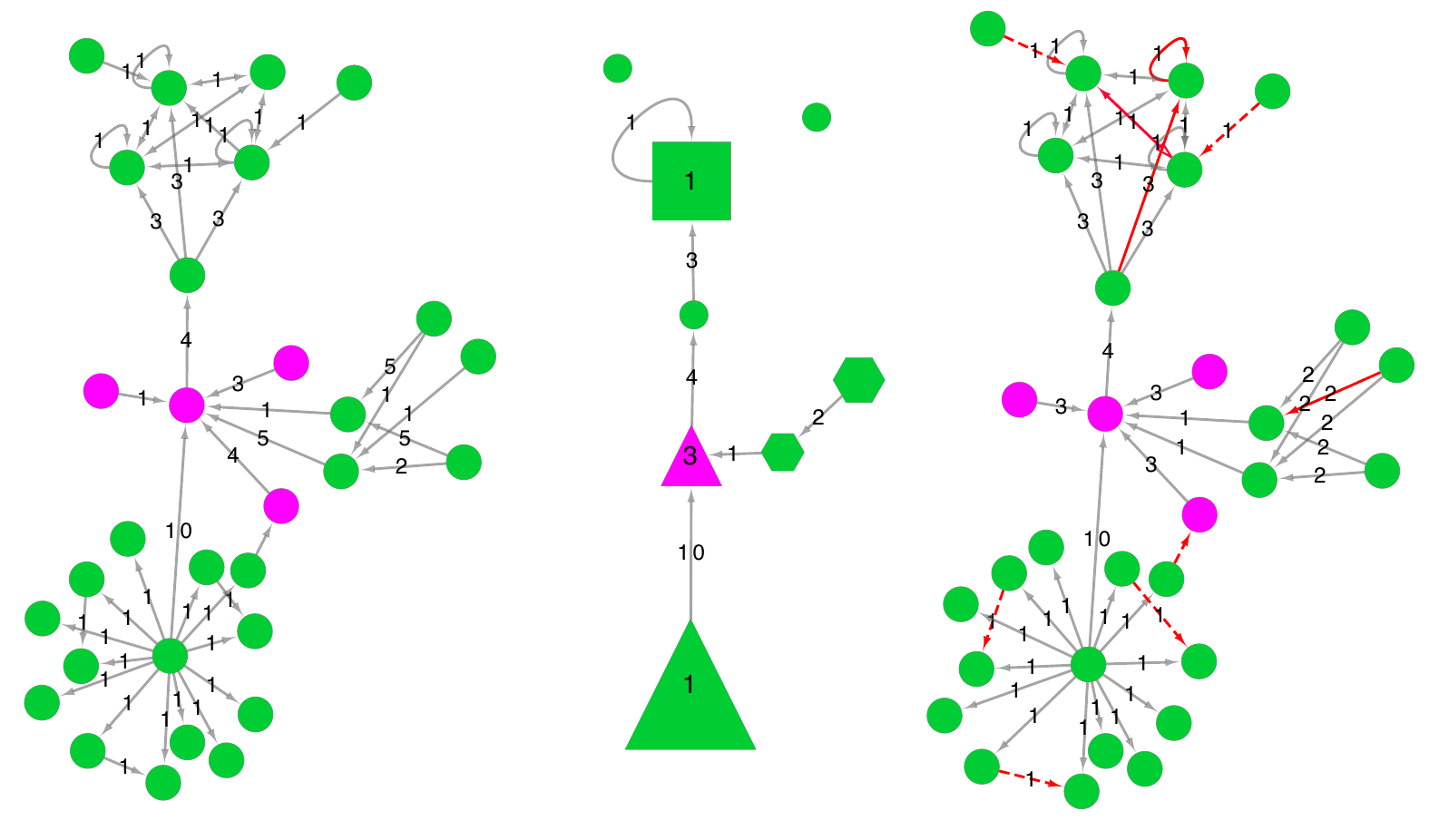}
	\vspace{-0.3in}
	\caption{(best in color) Ex. input graph (left), its summary/super-graph (middle), and the decompressed graph (right) w/ edge corrections in \textcolor{red}{red}, where dashed \protect\redlinedashed (solid \protect\redlinesolid)  are edges that need to be added %solid  for edges that need to be 
		(removed) for lossless reconstruction. See text for description of the scalars, node color, size, and shape. \label{fig:example}}
	\vspace{-0.65cm}
\end{wrapfigure}
Our goal is to output a small yet representative summary that facilitates the visualization, by which, improves the understanding of the overall structure of an input graph.
To this end, we model a \textit{summary graph (or super-graph)} as a collection of labeled super-nodes and weighted super-edges.
As illustrated in Fig. \ref{fig:example}, we merge structurally similar nodes 
of the same type/label (depicted by color)
into super-nodes  (size reflecting the number of constituent nodes). 
Super-nodes capture prevalent structural constructs found in real-world graphs, such as stars and cliques \cite{vog} %depicted by the glyph type 
(depicted by shape). A super-node is also marked with a scalar (i.e., weight), representative of the edge multiplicities among its nodes. 
A super-edge is placed between two super-nodes whose constituent nodes are sufficiently well-connected, and is also marked with a  scalar (i.e., weight) that best represents the edge multiplicities inbetween.

We aim to construct a small summary graph, which accurately reflects the input graph.
Here, succinctness and accuracy are in trade-off; the coarser the summary graph, the more information about the original graph is lost. 
We design a novel two-part \textit{lossless} encoding scheme, describing ($i$) the summary graph and ($ii$) the corrections required to reconstruct the input graph losslessly.
Treating the total number of encoding bits as a cost function, we design algorithms to find a summary with a small total cost. In summary, our main contributions are:

%We summarize our main contributions as follows:
\cbit
\item {\bf Summarizing LMDS-Graphs:~} \method is the first graph summarization method that can simultaneously handle node-\textbf{L}abels, edge \textbf{M}ultiplicities, edge \textbf{D}irections, and \textbf{S}elf-loops if any, as well as any combination thereof (Sec. \ref{04related}). LMDS-Graphs form the most general type of graphs, and frequently offer the most complete representation of domain specific data (e.g., accounting, social media, etc.).

\item {\bf Novel Super-graph Model:~} We build an interpretable summary graph consisting of super-nodes and super-edges. Super-nodes capture nodes with similar connectivity, e.g., within stars or cliques. A representative multiplicity is found for each set of summarized edges by a super-edge and within a super-node. (Sec. \ref{02modela})

\item {\bf Novel Two-part Lossless Encoding Scheme:~}
We introduce a novel scheme for encoding ($i$) the summary graph as well as ($ii$) the corrections for reconstructing the input graph losslessly. (Sec. \ref{02modelb})
%Based on the Minimum Description Length principle \cite{mdl_tutorial}, we employ the sum of bits as our cost function, 
%where the goal is to find a small summary graph (model) that requires few corrections (data given the model),  minimizing the total data description.

\item {\bf Efficient Search Algorithms:~} We propose a fast algorithm for searching groups of structurally-similar nodes to merge into super-nodes. 
Our candidate set generation 
avoids pairwise computations and allows multi-resolution summaries. (Sec. \ref{03algos}) 

%We obtain candidate sets via similarity-preserving hashing {\cite{Andoni2008}}, 
%avoiding pairwise computations.
%Sets of 
%increasing size are constructed by incrementally decreasing the similarity threshold, which allows multi-resolution summaries. (Sec. \ref{03algos})  

\ceit

%\reminder{Experiments ... kinds and high level results}

Extensive experiments on 12 real-world graphs with varying properties show the versatility of \method, which reveals insights that are otherwise hard to draw. % (Sec. \ref{ssec:qual}).
\method also achieves better compression with increasing allotted running time 
relative to the baselines on
comparable (i.e., non-LMDS) settings, and scales {near-linearly} with input size. (Sec. \ref{05eval}) % (Sec. \ref{ssec:quant}).

{\bf Reproducibility:} Source code for \method and all public-domain datasets are shared at {\url{https://anonymous.4open.science/r/4243006d-fede-49bd-b549-02387edf7ccd/}}.

\section{Related Work} \label{04related}

%\begin{wraptable}[!t]
\begin{wraptable}{l}{8.5cm}
	\vspace{-0.35in}
	\footnotesize
	\caption{ {\bf Our method matches \textit{all} specs}, while prior work
		miss one or more of the graph or summary properties. * depicts the cases that the method can be modified to handle, although details are not given in the corresponding paper. }
	\begin{center}
		\vspace{-0.25in}
		\begin{tabular}{ l| c| c | c | c ||c |c|c|}
			& \multicolumn{4}{c||}{\textbf{Input graph}} & \multicolumn{3}{c}{\textbf{Summary}} \\
			\toprule
			\diagbox{\textbf{Method}}{\textbf{Property}}  & \rotatebox{90}{{\bf L}abeled nodes}   & \rotatebox{90}{{\bf D}irected edges} & \rotatebox{90}{edge {\bf M}ultiplicity} & \rotatebox{90}{{\bf S}elf-loops} 
			& \rotatebox{90}{Super-graph} &  \rotatebox{90}{Lossless} & \rotatebox{90}{Multiresolution}  \\ 
			\midrule
			Khan \etal \cite{khan2015set}, SWeG \cite{shin2019sweg}  &  & & & & \textcolor{green}{\CheckmarkBold} & \textcolor{green}{\CheckmarkBold} & \\	
			%						Khan \etal \cite{khan2015set} & &   & & & \textcolor{green}{\CheckmarkBold} & \textcolor{green}{\CheckmarkBold} &  \\ 
			GraSS \cite{grass} &  &   & & & \textcolor{green}{\CheckmarkBold} & \textcolor{green}{\CheckmarkBold} & \textcolor{green}{\CheckmarkBold} \\ 						
			VoG \cite{vog}  &  &    & & & & \textcolor{green}{\CheckmarkBold} &  \\

			\hline	
			Riondato \etal\cite{quality_summary} &  & & \textcolor{green}{\CheckmarkBold}& \textcolor{green}{\CheckmarkBold}  & \textcolor{green}{\CheckmarkBold} & & \textcolor{green}{\CheckmarkBold} \\ %
			Toivonen \etal\cite{weighted_compression} &  &   & \textcolor{green}{\CheckmarkBold} & & \textcolor{green}{\CheckmarkBold} & & \textcolor{green}{\CheckmarkBold}   \\ 																			
			\hline
			CoSum \cite{zhu2016unsupervised}  &  \textcolor{green}{\CheckmarkBold} &    &  & & \textcolor{green}{\CheckmarkBold} &  &   \\

			Liu \etal \cite{liu2012approximate}  & \textcolor{green}{\CheckmarkBold} &  \textcolor{green}{\CheckmarkBold} & & & \textcolor{green}{\CheckmarkBold} & & \textcolor{green}{\CheckmarkBold}  \\								
			SNAP \cite{k_snap}  & \textcolor{green}{\CheckmarkBold} &  \textcolor{green}{\CheckmarkBold}$^\ast$ & && \textcolor{green}{\CheckmarkBold} & & \textcolor{green}{\CheckmarkBold}  \\ 
			
			Subdue \cite{subdue}  & \textcolor{green}{\CheckmarkBold} &  \textcolor{green}{\CheckmarkBold} &  & \textcolor{green}{\CheckmarkBold} & \textcolor{green}{\CheckmarkBold} &  & \textcolor{green}{\CheckmarkBold}\\ % 
			Navlakha \etal\cite{navlakha}  & \textcolor{green}{\CheckmarkBold}$^\ast$  &   \textcolor{green}{\CheckmarkBold}$^\ast$ & & &\textcolor{green}{\CheckmarkBold}  & \textcolor{green}{\CheckmarkBold} & \textcolor{green}{\CheckmarkBold} \\ 
			\midrule
			\method [this paper]  & \textcolor{green}{\CheckmarkBold} &  \textcolor{green}{\CheckmarkBold} & \textcolor{green}{\CheckmarkBold} &  \textcolor{green}{\CheckmarkBold} & \textcolor{green}{\CheckmarkBold} & \textcolor{green}{\CheckmarkBold} & \textcolor{green}{\CheckmarkBold} \\ \bottomrule				
		\end{tabular} 
		
	\end{center}
	\vspace{-0.35in}
	\label{tab:salesman}
\end{wraptable}

Graph \textit{summarization} and graph \textit{compression} techniques, while related, exhibit a key distinction. The former typically aims to simplify an input graph into a coarser one, while reflecting its prominent structure. On the other hand, the latter aims at reducing the storage requirements of a graph, often enabling speedy querying, while maintaining a certain level of query accuracy \cite{AdlerM01,BoldiV04,conf/wsdm/BuehrerC08,conf/icdm/KangF11,conf/cikm/LiakosPS14}. (See \cite{besta2018survey} for a recent survey.)
In this work we focus on graph summarization, with a goal to 
extract a simplified overview of key structural patterns within an input graph.
Table \ref{tab:salesman} provides a structured comparison of related work, with respect to the properties of the input graph as well as properties of the summary.

Most graph summarization techniques are designed for unlabeled, undirected, and simple graphs without edge multiplicities, weights, or self-loops \cite{shin2019sweg,khan2015set,grass,vog}. Closely-related are {graph-pooling} methods used within graph neural networks to gradually reduce the dimension of the layers; see. e.g. \cite{pooling2}.  
Riondato \etal\cite{quality_summary} and Toivonen \etal\cite{weighted_compression} are some of the few summarization methods that can accommodate weighted edges, but not labeled nodes or directed edges.
Among the methods that can handle graphs with multiple node labels,   
CoSum \cite{zhu2016unsupervised}, 
Liu \etal \cite{liu2012approximate},
and SNAP \cite{k_snap} build a coarser graph by only merging  the nodes of the same label into super-nodes.
Differently, Subdue \cite{subdue} replaces frequent sub-graphs that potentially contain different labels with a super-node, which makes the interpretation of the summary graph harder.

Closest to our work is the approach by Navlakha \etal\cite{navlakha}, which iteratively merges nodes into super-nodes as long as the description length of the input graph decreases. Thanks to its simple model and algorithm, it can be modified to handle labeled graphs, specifically by restricting the node merges to same-label nodes. However, its model is unable, nor is it trivial to modify, to accommodate edge weights/multiplicities.
All in all, there is {\em no existing work} that can summarize labeled multi-graphs -- with labeled nodes, directed and multi-edges and self-loops. (See \cite{journals/csur/LiuSDK18} for an extended survey and Table 1 therein.)  
While our \method is the first of its kind, it is \textit{versatile} in that it can also accommodate graphs with any combination of those properties.

Besides input graph properties, prior work can also be classified w.r.t. the properties of the summary. Here, we focus on summaries where the output is itself a (coarser) graph, called the summary or super graph.
%In principle, summary graph consists of super-nodes and super-edges that reflect the high-level sub-structures and relationships. 
VoG \cite{vog} is an exception in Table  \ref{tab:salesman}; it identifies key sub-structures (stars, (near)cliques, etc.) however does not provide any super-edges, i.e., its summary graph is disconnected.
Second, the summary may be lossy; including only the coarse summary graph \cite{quality_summary,weighted_compression,zhu2016unsupervised,liu2012approximate,k_snap,subdue},
or lossless; consisting of both the summary and the corrections necessary to fully reconstruct the input graph \cite{shin2019sweg,khan2015set,grass,vog,navlakha}.
Finally, a desired characteristic of a summary is multi-granularity; where the  coarseness or  resolution of the summary graph can be adjusted on demand \cite{grass,quality_summary,weighted_compression,liu2012approximate,k_snap,subdue,navlakha},  via appropriately altering some of the model parameters.
Notably, \method exhibits all of these three properties: super graph output, lossless and multi-resolution summary.

\section{Graph Summary Design and Encoding}
\label{02model}

\subsection{Summary and Decompression}
\label{02modela}
Given a directed graph $\mathcal{G} = \{ \mathcal{V}, \mathcal{E}, \mathcal{T} \}$ with edge multiplicities $m(e)\in \mathbb{N}, \forall e\in \mathcal{E}$, node labels/types $\ell(u)\in \mathcal{T}, \forall u\in \mathcal{V}$, and  self-loops, we define the \emph{summary} and \emph{decompressed} graphs as follows.

\vspace{-0.1in}
\subsubsection{\bf 3.1.1 Summary graph (or super-graph) model}
Let $\mathcal{G}_\mathrm{s} = \{ \mathcal{V}_\mathrm{s}, \mathcal{E}_\mathrm{s} \}$ be  the sets of super-nodes and directed super-edges that define the summary graph topology. Each super-node $v \in \mathcal{V}_\mathrm{s}$ is annotated by four components: ($i$) its label $\ell(v)$ (depicted by color), ($ii$) the number $|\mathcal{S}_v|$, $\mathcal{S}_v \subset \mathcal{V}$,  of nodes it contains (depicted by size), ($iii$) the \emph{glyph} $\mu(v)\in\mathcal{M}$ it represents (depicted by shape), and ($iv$) the \emph{representative} multiplicity $m(v)$ of the edges it summarizes (depicted as a scalar inside the glyph). For each super-edge $e \in \mathcal{E}_s$, we let $m(e)$ be the \textit{representative} multiplicity of the edges it captures, depicted as a scalar on the super-edge. 
(We describe how to find the ``representative'' multiplicity of a set of edges in Sec. \ref{find_rep}.)

Fig. \ref{fig:example} (left) and (middle) respectively depict an example input graph and its corresponding summary graph. Apart from unmerged simple nodes that are depicted as plain circles, the set $\mathcal{M}$ of possible glyphs that \method supports contains: 1) {\tt Clique} (square), 2) {\tt In-star} (triangle), 3) {\tt Out-star} (triangle), and 4) {\tt Disconnected} set (hexagon). Such structures are commonly found in real-word graphs \cite{vog}. For instance, a clique can represent a tightly-knit group of friends in a social network, while an out-star can capture spam-like activity in an email or call network. Moreover, %mapping 
using glyphs %to appropriate shapes 
has been shown to yield easily interpretable visualizations \cite{motif}. 

\vspace{-0.1in}
\subsubsection{\bf 3.1.2 Decompression}\label{dec}
The summary graph $\mathcal{G}_\mathrm{s}$ decompresses \emph{uniquely} and \emph{unambiguously} into 
$\mathcal{G}^\prime = \mathrm{dec}(\mathcal{G}_\mathrm{s}) = \{\mathcal{V}, \mathcal{E}^\prime \}$
%where the edge set
%\begin{align}
%\mathcal{E}^\prime = \mathcal{E}^\prime_\mathrm{GL} \cup \mathcal{E}^\prime_\mathrm{SE} \cup \mathcal{E}^\prime_\mathrm{SL}
%\end{align}
according to simple and intuitive rules (e.g., Fig. \ref{fig:example} (right)).  First, every super-node expands to the set of nodes it contains, all of which also inherit the super-node's label. The nodes are then connected according to the super-node's glyph: for out(in)-stars a node defined as the hub points to (is pointed by) all other nodes, for cliques all possible directed edges are added between the nodes, and for disconnected sets no edges are added. Moreover, a super-node self-loop expands to self-loops on every node it contains. On the other hand, super-edges expand to sets of edges that have the same direction. If the involved source/target glyphs  are not stars, all the nodes contained in the source glyph point to all the nodes contained in the target glyph. For stars, the expanded incoming and out-going super-edges are only connected to the star's hub. Finally, all expanded edges inherit their corresponding ``parent''  super-node's or super-edge's representative multiplicity. 

Apart from enabling a clear interpretation of a given summary, the decompression rules help quantify how well the summary represents the original graph. For example, the \textcolor{purple}{pink} triangle with representative multiplicity 3 in Fig. \ref{fig:example} (middle) expands to an in-star with all edges having multiplicity 3 as shown in Fig. \ref{fig:example} (right). While the topology is perfectly captured (\textcolor{purple}{pink} nodes form a perfect in-star), the expanded multiplicities are not always equal to the original ones. On the other hand, expanding the \textcolor{green}{green} triangle perfectly captures the edge multiplicities (all are 1), but only approximates the topology, as the original \textcolor{green}{green} subgraph also contains some edges between the spokes of the hub node. %Positive corrections (edges to be added to $\mathcal{G}^\prime$), and negative ones (to be removed), are represented in Fig. 1 with dashed and solid \textcolor{red}{red} lines respectively.     

\subsection{Model Encoding}\label{encoding}
\label{02modelb}

Following the two-part Minimum Description Length paradigm \cite{mdl_tutorial}, we aim to identify a summary graph $\mathcal{G}_\mathrm{s}$ that minimizes the total description cost of the full graph, that is,

\vspace{-0.15in}
%\small
\begin{align}\label{prob}
\hspace{-0.1in} \mathcal{G}^\ast_\mathrm{s} :=& \arg\underset{\mathcal{G}\mathrm{s}}{\min} \;\;\; L(\mathcal{G}_\mathrm{s}) \; +\; L(\mathcal{G}|\mathcal{G}^\prime), ~~\mathrm{s.t.}~~\;\; \mathcal{G}^\prime = \mathrm{dec}(\mathcal{G}_\mathrm{s})
\end{align}
%\normalsize
where $L(\mathcal{G}_\mathrm{s})$ measures the number of bits required to encode the summary graph, and $L(\mathcal{G}|\mathcal{G}^\prime)$ the bits needed to encode the corrections (or extra-information) for reconstructing the original graph $\mathcal{G}$ from the (uniquely and unambiguously) decompressed $\mathcal{G}^\prime$. These costs can be quantified as follows.

\vspace{-0.1in}
\subsubsection{\bf 3.2.1 Encoding the summary graph}
We first encode the size of the summary graph $L_\mathbb{N}(|\mathcal{V}_\mathrm{s}|)$, and the number of labels $L_\mathbb{N}(|\mathcal{T}|)$.\footnote{\label{note1}$L_\mathbb{N}(k)=2\log k +1$ bits are required to encode an arbitrarily large natural number $k$, using the variable-length prefix-free encoding; see, Ex. 2.4 in \cite{mdl_tutorial}. }
For each super-node,  $\log_2|\mathcal{T}|$ bits are used to record its label, $\log_2|\mathcal{M}|$ for its glyph, $L_\mathbb{N}(|\mathcal{S}_v|)$ for its size, $\mathcal{L}_\mathbb{N}(m(v))$ for the within-glyph representative multiplicity, $\log_2(|\mathcal{V}_\mathrm{s}|)$ for the number of super-nodes in $\mathcal{G}_\mathrm{s}$ that it points to, and $\log_2{|\mathcal{V}_\mathrm{s}| \choose |\mathcal{N}(v)|}$ to identify the specific set of super-nodes it points to, where $\mathcal{N}(v)$ denotes the set of direct (out)neighbors.  
For each super-edge, $L_\mathbb{N}(m(e))$ bits are used for the representative multiplicity. In total, the number of bits required to encode a summary graph is given as
%\small
\begin{align}
L(\mathcal{G}_\mathrm{s}) = & L_\mathbb{N}(|\mathcal{V}_\mathrm{s}|) + L_\mathbb{N}(|\mathcal{T}|)+ \sum_{v\in\mathcal{V}_\mathrm{s}}L_\mathrm{SNODE}(v)\;, \text{ where}
\end{align} 
\vspace{-0.15in}
\begin{equation}
%\resizebox{.8\hsize}{!}{
L_\mathrm{SNODE}(v) = \log_2|\mathcal{T}| + \log_2|\mathcal{M}| + L_\mathbb{N}(|\mathcal{S}_v|)+ \mathcal{L}_\mathbb{N}(m(v)) 
+ \log_2(|\mathcal{V}_\mathrm{s}|) +   \log_2{|\mathcal{V}_\mathrm{s}| \choose |\mathcal{N}(v)|} + \sum\limits_{z\in\mathcal{N}(v)} L_\mathbb{N}(m(v,z)) 
%}
\vspace{-0.15in}
\end{equation}

%\normalsize

\subsubsection{\bf 3.2.2 Encoding the corrections}\label{corrections}
For the overall cost for corrections, we first compute the number of bits used to correct the \textit{topology} of the expanded (i.e., decompressed) graph, followed by the number of bits needed to represent the true \textit{multiplicities}.  Regarding the topology, we first  map the expanded nodes back to the original node-set $\mathcal{V}$. 
This costs  $L_\mathrm{MAP}(v) = \log_2{|\mathcal{V}| \choose |\mathcal{S}_v|} + \mathds{1}_{\{\mu(v)=\mathrm{STAR}\}}\log_2|\mathcal{S}_v|$ bits per super-node $v$ ({the latter term identifying the hub of a star}). 
Subsequently, we have two types of edge corrections: Either adding edges that exist in the original graph but not in the expanded graph (i.e., positive corrections)
%$$\mathcal{E}_\mathrm{COR}^+(v,z) = \mathcal{E}_{\mathcal{S}_v,\mathcal{S}_z}\setminus \mathcal{E}^\prime_{\mathcal{S}_v,\mathcal{S}_z}(\mu(v),\mu(z)), $$
or removing edges from the expanded graph because they do not exist in the original graph (i.e., negative corrections). 

These costs are compactly encoded for every expanded super-edge and every expanded super-node (glyph), using the binomial encoding 
$L(\mathcal{E}_\mathrm{COR}) = L_\mathbb{N}(|\mathcal{E}_\mathrm{COR}|) + \log_2{ |\mathcal{E}^{\max}_\mathrm{COR}| \choose |\mathcal{E}_\mathrm{COR}| }$, where $\mathcal{E}_\mathrm{COR}$ denotes the possible set of corrections (positive or negative), and $\mathcal{E}^{\max}_\mathrm{COR}$ the largest set that $\mathcal{E}_\mathrm{COR}$ can possibly be. For example, for positive edge corrections in a disconnected set, we have  $\mathcal{E}^{\max}_\mathrm{COR} = \mathcal{S}_v \times \mathcal{S}_v$, and similarly for negative edge corrections in a clique. For super-edges, corrections are computed according to the decompression rules (see Sec. \ref{dec}). For the (few) edges in the original graph between super-nodes $v$ and $z$ that are not represented by a super-edge, the corrections are always positive, and $\mathcal{E}^{\max}_\mathrm{COR} = \mathcal{S}_v \times \mathcal{S}_z$.

The binomial encoding arises from using the uniform code over all the lexicographically ordered \textit{subsets} of possible corrections. An alternative to this, as suggested in \cite{grass} and \cite{quality_summary}, would be to encode each correction \emph{individually} using an optimal prefix code. Then, interpreting  $p = |\mathcal{E}_\mathrm{COR}|/|\mathcal{E}^{\max}_\mathrm{COR}| $ as the ``probability'' of each correction, we would need  
%\begin{align}
%L_\mathrm{ENTR} &= \sum\limits_{u_1\in\mathcal{S}_v}\sum\limits_{u_2\in\mathcal{S}_z}\bigg(\log_2\frac{1}{p}\cdot\mathds{1}_{\{ (u_1,u_2)\in\mathcal{E}_{\mathcal{S}_v,\mathcal{S}_z} \}} \\
%&+ \log_2\frac{1}{1-p}\cdot\mathds{1}_{\{ (u_1,u_2)\notin\mathcal{E}_{\mathcal{S}_v,\mathcal{S}_z} \}} \bigg) \\
%&= n\cdot H(\mathrm{Ber}[p])
%\end{align}
$L_\mathrm{ENTR} = H(\mathrm{Ber}[p]) \cdot |\mathcal{E}^{\max}_\mathrm{COR}|$
bits, where $H(\cdot)$ is the Shannon entropy, and $\mathrm{Ber}[p]$ is a Bernoulli with parameter $p$. Denoting $|\mathcal{E}_\mathrm{COR}|=n^\prime$ and $|\mathcal{E}^{\max}_\mathrm{COR}|=n$, we can show that our binomial encoding is more efficient. 
%\newtheorem{theorem}{Theorem}
%\vspace{0.1in}
\begin{theorem}
	It holds that 
	%	\small 
	%\vspace{-0.2cm}
%	\begin{equation}\label{eq2}
$
	L_\mathrm{ENTR} \geq \log_2{ n \choose n^\prime}\;\;,
	\;\;\text{for all $n > n^\prime>0$.}
	$
%	\vspace{-0.4cm} 
%	\end{equation}
	\label{prop1}
	%	\normalsize
\end{theorem}
\vspace{-0.15in}
\begin{proof}
	See Appendix \ref{ssec:thm1}.
\end{proof}
%\vspace{0.1in}
Theorem \ref{prop1} establishes that the binomial encoding always gives a more compact measure of information required for corrections. 
Having corrected the edge topology, we compute the cost of correcting the edge multiplicities. Since any edge $e$ not included in a glyph or super-edge does not have a representative multiplicity, its multiplicity correction is encoded by $L_\mathbb{N}(m(e))$, encoding its true value. The reason for using $L_\mathbb{N}(\cdot)$ to encode multiplicities is the fact that, for most real graphs, multiplicities follow a power-law distribution. Since the vast majority has small values, $L_\mathbb{N}(\cdot)$ will generally be a more ``compact'' encoding compared to a uniform code based on the maximum multiplicities. For expanded super-nodes and super-edges with representative multiplicity $m$, we obtain the cost of correcting the multiplicities as
\begin{align}
L_\mathrm{DIFF}(\mathcal{E}_{\text{sup}},m) = \sum_{e\in\mathcal{E}_{\text{sup}}}\ell_\mathrm{diff}(m(e),m),   
\label{ldif}
\end{align}
%\normalsize
where $\mathcal{E}_{\text{sup}}$ in this context is the set of all edges contained in said super-node or super-edge, and
%\small
\begin{align} 	
\ell_\mathrm{diff}(m^\prime,m)=\left\{ \begin{array}{cc}
1\;,~&~m= m^\prime\\
2\log_2(|m-m^\prime|) + 3 \;,& m\neq m^\prime 
\end{array} \right., 
\end{align}
%\normalsize
bits are  needed to encode the \emph{difference} between a true multiplicity $m^\prime$ and its representative $m$. Note that, since $L_\mathbb{N}(\cdot)$ only holds for natural numbers (see footnote\footnoteref{note1}), 
one extra bit is required to indicate whether the difference is 0, and one more 
for the sign of the difference.

\section{Graph Summary Search}
\label{03algos}

The discrete optimization problem in \eqref{prob} has a very large set of feasible solutions, and needs to be approximated efficiently. Towards this goal, we follow a two-step process, where we first generate a list of (possibly overlapping) groups of nodes, which we term \emph{candidate} node-sets (see Sec. \ref{candidates}), and then decide which ones to merge into super-nodes. These candidates have varying size and quality (i.e., structural-similarity). Larger candidates with low quality compress the graph more (reduced $L(\mathcal{G}_\mathrm{s})$), but also typically require more corrections (increased $L(\mathcal{G}|\mathcal{G}^\prime)$). Clearly, the best candidates have both high quality and large size. For this reason, we first sort the candidate sets in descending order with respect to the product of their size and quality. We then process the sorted list from top to bottom, and merge the candidate sets into super-nodes, updating the summary graph accordingly (see Sec. \ref{merging}). To ensure the quality of summarization, we only monitor the overall total cost, and only commit to a given candidate if  $\Delta_\mathrm{cost} = \mathrm{Cost\_After} - \mathrm{Cost\_Before} <0$. This offers two benefits: (1) We avoid the cumbersome process of merging nodes in pairs (i.e. two at a time) and instead merge \textit{in groups}, and (2) We achieve ability to summarize at \textit{multiple resolutions}. The overview is given in Algorithm \ref{alg5}.
\vspace{-0.1in} 
\begin{algorithm}
	\KwIn{directed labeled multi-graph $\mathcal{G}$ }
	
	Construct candidate node-sets (Sec. 4.1)\;	
	Sort candidates w.r.t. $(\mathrm{size}\times\mathrm{quality})$\;
	\For{ every candidate set in list}
	{	
		Merge unmarked nodes in set and decide glyph (Sec. 4.2.1)\;
		Decide super-edges (Sec. 4.2.2) \; 	
		Compute representative multiplicities (Sec. 4.2.3)\;
		Mark candidate node-set as merged\;	
		\If{$\Delta_\mathrm{cost}<0$}
		{
			Commit to merged super-node and its super-edges\;
		}  		
	}
	\textbf{Return} summary graph $\mathcal{G}_\mathrm{s}$ \;
	\caption{{\small{\method: Summarizing Labeled Multi-Graphs}}}
	\label{alg5}
\end{algorithm}
\vspace{-0.2in}

\subsection{Candidate Set Generation}
\label{candidates}

\subsubsection{\bf 4.1.1 Measuring candidate quality}

\begin{wrapfigure}{r}{0.21\textwidth}
	\centering
	\vspace{-0.35in}
	\includegraphics[width=35mm]{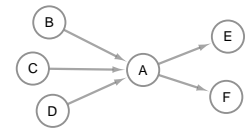}\\
	\vspace{-1.1 in}
	\caption{Directed Jaccard similarity treats incoming (B, C, D) and outgoing (E, F) neighbors separately.}
	\label{fig:jac}
	\vspace{-0.35in}
\end{wrapfigure}
To quantify a candidate set's quality, we first need to define a proper metric of structural node similarity. For undirected graphs, the Jaccard similarity between two nodes $v$ and $v^\prime$ is given as 
%\small 
%\begin{align}\label{jack}
$
J^U(v,v^\prime) = \frac{ | \mathcal{N}^U(v)\cap \mathcal{N}^U(v^\prime)| }{ | \mathcal{N}^U(v)\cup \mathcal{N}^U(v^\prime)| }$,
%\end{align}
%\normalsize
and simply measures the proportion of common neighbors that they share. Na\"ively using $J^U(\cdot,\cdot)$ on directed graphs is straightforward by ignoring the directions of the edges, however, it may yield misleading results by often over-estimating the true node similarity.

Consider the simple example depicted in Fig. \ref{fig:jac}, where any pair in the set $\{B,C,D,E,F\}$ has undirected similarity $1$, while the fact that $\{B,C,D\}$ are  incoming and $\{E,F\}$ are outgoing means that the set is far from an ideal candidate. 
To mitigate such inconsistencies, we introduce the following extension of Jaccard that may also accommodate directed graphs, by taking into account the similarity of both $I$ncoming and $O$utgoing edges.

\begin{definition}
	The \emph{directed Jaccard} similarity between any two pair of nodes $v,v^\prime$ of a directed graph is given as 
	%	\small
	\begin{align}\label{jack}
	J^D(v,v^\prime) = \frac{ | \mathcal{N}^I(v)\cap \mathcal{N}^I(v^\prime)| + | \mathcal{N}^O(v)\cap \mathcal{N}^O(v^\prime)| }{| \mathcal{N}^I(v)\cup \mathcal{N}^I(v^\prime)| + | \mathcal{N}^O(v)\cup \mathcal{N}^O(v^\prime)| }
	\end{align}
\end{definition}
%\normalsize
First, it can easily be observed that for undirected graphs, $J^D(v,v^\prime)$ $=J^U(v,v^\prime)$, since $\mathcal{N}^I(v)=\mathcal{N}^O(v)=\mathcal{N}^U(v)$. Note however, that in our example, $J^D(v,v^\prime)$ becomes $0$ for all cross-pairs between $\{B,C,D\}$ and $\{E,F\}$, effectively creating two separate groups. In general for directed graphs, $J^D(v,v^\prime)$ will be more ``informed'' than $J^U(v,v^\prime)$, typically yielding lower similarity scores.

Having defined our similarity, we measure a candidate's quality according to the minimum directed Jaccard  among all pairs of nodes it contains. Thus, given the following definition,

\begin{definition}
	Any set $\mathcal{C}\subseteq \mathcal{V}$, is \emph{$t-$bounded} if $\;\;\; J^D(v,v^\prime)\geq t \;\; ~\forall (v,v^\prime)\in\mathcal{C}\times\mathcal{C} \;\;.$ 
\end{definition}
We use the $t-$bounded-ness of a candidate to serve as a pessimistic valuation of its quality. In addition, given that we are interested in a \emph{collection} of candidate sets, we would like the sets to be \emph{non-redundant} defined as follows.

\begin{definition}\label{non_red}
	Let $\mathcal{C}_\mathrm{S}$ be a collection of candidate sets, each one accompanied by a bound $t$.  We call $\mathcal{C}_\mathrm{S}$ \emph{non}-\emph{redundant}, if for any $\mathcal{C}\in\mathcal{C}_\mathrm{S}$ that is $t-$bounded, there exists \textit{no} $t^\prime-$bounded $\mathcal{C}^\prime\in\mathcal{C}_\mathrm{S}$, such that $t^\prime \geq t$ and $\mathcal{C}\subset \mathcal{C}^\prime$.
\end{definition}
Simply put, non-redundancy ensures that none of the candidate sets is a strict subset of another set of higher or equal quality.

\vspace{-.1in}
\subsubsection{\bf 4.1.2 Incremental LSH} To group nodes according to their similarity, we first utilize Locality Sensitive Hashing (LSH) \cite{Andoni2008}. Specifically for every node $v$, we generate a set of $r$ \emph{minhash} signatures
%\small
\begin{align}\label{min_h}
h_j(v) := \underset{z\in \mathcal{N}^D(v)}{\min} \; f_j(z)~~\;\;\; \forall j=1,\ldots,r 
\end{align}
%\normalsize
where $f_j$'s are independent and uniform hash functions (see, e.g., \cite{Andoni2008} for implementation details), and $\mathcal{N}^D(v) := \mathcal{N}^I(v) \| \mathcal{N}^O(v)$ is the concatenated adjacency list of node $v$ that includes all incoming and outgoing neighbors separately. It can then be shown that  
%\small
%\begin{align}\label{min_h_prop}
%
$
\Pr\big\{ h_j(v) = h_j(v^\prime)\big\} = 
%\frac{|\mathcal{N}^D(v)\cap\mathcal{N}^D(v^\prime)|}{|\mathcal{N}^D(v)\cup\mathcal{N}^D(v^\prime)|} = 
J^D(v,v^\prime) %\;,
%\end{align}
$;
%\normalsize
that is, two nodes share a minhash signature with probability proportional to their directed Jaccard similarity. % \cite{Andoni2008}.

Since the $r$ hash functions are independent, it follows that
%\small
%\begin{align}
%\Pr\big\{\mathbf{h}(v) = \mathbf{h}(v^\prime)\big\} &= \Pr\big\{ h_j(v) = h_j(v^\prime)~~\forall j=1,\ldots,r\big\} 
%= \left(J^D(v,v^\prime)\right)^r \label{mult}
$\Pr\big\{\mathbf{h}(v) = \mathbf{h}(v^\prime)\big\} 
= \left(J^D(v,v^\prime)\right)^r 
%\label{mult}
%\end{align}
$, 
%\normalsize
where $\mathbf{h}(v) := \left[ h_1(v),\ldots,h_r(v) \right]^T$ is the $r-$length minhash signature vector of node $v$. If the nodes are hashed into buckets according to their $r$ minhash signatures, %Eq. \eqref{mult} is
the equality gives
 the probability that two nodes hash into the same bucket. By collecting $b$ hash-tables corresponding to $b$ bands of $r$ minhash signatures, the probability that $v$ and $v^\prime$ hash to the same bucket at least once is %given by 
%\small
\begin{align}\label{at_least_one}
\Pr\big\{\mathbf{h}_i(v) = \mathbf{h}_i(v^\prime)~ \exists i=1,\ldots,b\big\} = 1 - \left(1 - \left(J^D(v,v^\prime)\right)^r\right)^b
\end{align}
%\normalsize

Interestingly, for sufficiently large $r$ and $b$, the RHS expression in Eq. \eqref{at_least_one} when viewed as a function of $\left(J^D(v,v^\prime)\right)$ approximates a step function around the threshold
$t=\left(\frac{1}{b}\right)^\frac{1}{r} \in (0, 1]$, meaning that with high probability $v$ and $v^\prime$ will belong in a $t-$bounded set. 
%\small
%\begin{align} 	
%\Pr\{\mathbf{h}_i(v) = \mathbf{h}_i(v^\prime)~~ \exists i=1,\ldots,b\} \approx \left\{ \begin{array}{cc}
%1 ,~&~ J(v,v^\prime) \geq t \\
%0,~&~ J(v,v^\prime) < t
%\end{array} \right., 
%\end{align}
%\normalsize
To avoid repeating the entire process for different values of $b$, we incrementally generate and add more bands of minhash node signatures, that in turn hash nodes into new buckets. The new buckets are then merged with any overlapping existing buckets, gradually coalescing  into larger clusters that are \emph{approximately}  
$t-$bounded, with $t=\left(\frac{1}{b}\right)^\frac{1}{r}$ decreasing as $b$ increases. This is exactly how we  obtain larger candidate sets, albeit of lower quality, incrementally by the addition of new bands.

\vspace{-.1in}
\subsubsection{\bf 4.1.3 Filtered LSH} While the incremental LSH described in the previous section efficiently guides the process of forming candidate sets, merged buckets are not guaranteed to be $t-$bounded due to the false alarm probability. For this purpose, we maintain an undirected similarity graph $\mathcal{G}_\mathrm{sim}$, where an edge $(v,v^\prime)$ is \emph{guaranteed} to appear if and only if 
$J^D(v,v^\prime)\geq t$. Intuitively, $\mathcal{G}_\mathrm{sim}$ serves as a data structure where large $t-$bounded candidates appear as maximal cliques. As new LSH buckets appear and clusters are updated, we compute $J^D(v,v^\prime)$ for newly coalesced pairs of nodes $(v,v^\prime)$, and add the latter as an edge to $\mathcal{G}_\mathrm{sim}$ if $J^D(v,v^\prime)\geq t$. If the threshold is not satisfied, the computed value $J^D(v,v^\prime)$ is not discarded, but cached into a max-heap since it may satisfy a lower $t$ in one of the subsequent iterations as $b$ is increased. To further reduce the complexity of verifying pair similarities, we rely on the Jaccard upper bound $J^D(v,v^\prime)\leq \min(d,d^\prime)/\max(d,d^\prime)$, where $d=|\mathcal{N}^D(v)|$. If the upper bound of a pair $(v,v^\prime)$ is smaller than the lowest threshold $t_{\min}=\left(\frac{1}{b_{\max}}\right)^\frac{1}{r}$ that will ever be required, we can safely ignore $(v,v^\prime)$ since it will never form a valid edge. Leveraging this property, we maintain nodes sorted in every cluster according to their in-plus-out-degrees $d$, and for a given $v^\prime$ in the array, we only consider similarities with $v$'s in the window $ \left \lfloor{ d_v t_{\min}}\right \rfloor \leq d_{v^\prime} \leq  \left \lceil{d_v/t_{\min}}\right \rceil $.

As mentioned earlier, candidate sets are collected as maximal cliques in $\mathcal{G}_\mathrm{sim}$. To ensure that the set of candidates is \emph{non-redundant} (cf. Def.n \ref{non_red}), we maintain for every node the size of the maximum clique that it has been found to belong in. Every time a new clique is discovered, we update the maximum-sizes for all the nodes it contains using the clique's size. As new edges are added to $\mathcal{G}_\mathrm{sim}$, we examine every node for newly emerged cliques, and we rely on the heuristic in \cite{clique} to prune the search by avoiding the evaluation of cliques that cannot exceed the size of the previously-found maximum clique. This way, we achieve two goals: 1) We ensure that the new cliques discovered at every iteration are not subsets of (or equal to) any previous clique, and
2) We reduce computation by early-stopping of the clique discovery process for most nodes.

\subsection{Merging Candidates: Glyphs, Super-edges, Multiplicities}
\label{merging}
Every time a candidate set $\mathcal{C}$ is tested, we deploy subroutines that efficiently update the summary graph, by making decisions regarding (1) the glyph that will be assigned to the merged set of nodes, (2) super-edges that emerge (or disappear) due to the merging, as well as (3) representative multiplicities for the set of edges summarized by the glyph and its super-edges.

\vspace{-.1in}
\subsubsection{\bf 4.2.1 Glyph decision rules}\label{dec_glyph} 
To preserve super-node label homogeneity, a candidate set that contains nodes of different labels is first split into same-label subsets. Each subset is merged into a separate super-node using the procedure described below. Hereafter, the term candidate set refers to such a label-homogeneous subset. 

For the glyph decision, we first identify the number of directed edges $E_\mathcal{C}$ that are included in the  subgraph induced on nodes that corresponds to $\mathcal{C}$ in the candidate set. Consequently, if $E_\mathcal{C}\geq |\mathcal{C}|(|\mathcal{C}|-1)/2$, i.e., at least half of all possible directed edges are present, then we decide {\tt Clique} since it most likely is the best glyph in terms of number of edge corrections. For sparsely-edged candidate sets that do not pass the clique threshold, we proceed to test for the presence of stars. If there is a suitable {out-}/{in-star} present in $\mathcal{C}$, then its hub will be the highest out/in-degree node in $\mathcal{C}$. We therefore compute $d^I_{\max}$ and $d^O_{\max}$, that is, the largest in and out degrees in the induced subgraph, and compute the following proxy correction cost for encoding an in-star 
%\small
\begin{equation}\label{proxy}
\mathrm{Cost}_\mathrm{IN} = (|\mathcal{C}| - 1  - d^I_{\max}) + (E_\mathcal{C} - d^I_{\max}) \;,
\end{equation}
%\normalsize
and similarly $\mathrm{Cost}_\mathrm{OUT}$ for out-star using $d^O_{\max}$. Intuitively, the first term of Eq. \eqref{proxy} is the number of edges that will have to be removed from the full decompressed star, while the second part is the number of edges that cannot be ``explained'' by the star and will have to be added. We then compare $\mathrm{Cost}_\mathrm{IN}$ and $\mathrm{Cost}_\mathrm{OUT}$ with $E_\mathcal{C}$, i.e., the number of edges that will have to be added if we decide that $\mathcal{C}$ is a {disconnected} set. If only $\mathrm{Cost}_\mathrm{IN}$ (or only $\mathrm{Cost}_\mathrm{OUT}$) is smaller than $E_\mathcal{C}$, then we decide {\tt In-star} (or {\tt Out-star}). If both  $\mathrm{Cost}_\mathrm{IN}$ and $\mathrm{Cost}_\mathrm{OUT}$ are smaller than $E_\mathcal{C}$, then we choose the smallest of the two. Finally, if neither $\mathrm{Cost}_\mathrm{IN}$ nor $\mathrm{Cost}_\mathrm{OUT}$ are smaller than $E_\mathcal{C}$, we decide that $\mathcal{C}$ is a {\tt Disconnected} set.

\vspace{-.1in}
\subsubsection{\bf 4.2.2 Super-edge decision rule}\label{dec_se} Having decided the glyph of $\mathcal{C}$, we merge any outgoing and incoming edges and/or super-edges into ``bundles'' of edges and their corresponding multiplicities. We then obtain the topology-based correction costs of merging or not merging each bundle into a super-edge (recall Sec. \ref{corrections}). 
Furthermore, we need to extract a representative multiplicity (see below) and compute the cost of encoding the corrections as the difference between the true 
%edge multiplicity 
and the representative multiplicity (recall Eq. \eqref{ldif}). If the total cost (topology and multiplicities) of representing each bundle of edges with a super-edge is lower than the cost of \emph{not} representing it, then the corresponding super-edge (and its representative multiplicity) is added to the summary. % graph.

%Show comparissons of how mean, median and mode fail to capture prepsesentative multiplicity. Also show how the total representation cost reduces.
%\begin{align*}
%m^\ast = \arg\min_m L_\mathrm{DIFF}(\mathcal{E},m)   
%\end{align*}

\vspace{-.1in}
\subsubsection{\bf 4.2.3 Finding representative multiplicities}\label{find_rep} For every newly-formed super-node as well as each potential super-edge, we find the representative multiplicity $m^\ast$ as

\vspace{-0.2in}
\begin{align}
m^\ast := \arg\min_m \;\;\; L_\mathrm{DIFF}(\mathcal{E}_{\text{sup}},m),
\end{align}
\vspace{-0.1in}

\noindent
where $L_\mathrm{DIFF}(\mathcal{E}_{\text{sup}},m)$ is defined in Eq. \eqref{ldif}, and $\mathcal{E}_{\text{sup}}$ is the set of all edges contained in a given super-node or bundled by a super-edge.
Although this 1D-optimization problem is not convex, we find that the dichotomous search algorithm \cite{chong2004introduction} 
finds the optimal solution in most cases, and runs in $\mathcal{O}(|\mathcal{E}_{\text{sup}}|\log_2R)$ time, where $R = \max_{e \in \mathcal{E}_{\text{sup}}} m(e)- \min_{e \in \mathcal{E}_{\text{sup}}} m(e)$, i.e., the dynamic range of multiplicities.

\vspace{-0.05in}
\subsection{Complexity Analysis}
\label{ssec:complexity}

{\bf Time:} First we analyze the complexity of the candidate generation algorithm. For a given LSH band of $r$ rows,  generating the minhash signatures is $\mathcal{O}(|\mathcal{E}|r)$, where $\mathcal{E}$ is the set of all directed edges in $\mathcal{G}$.
Further, filtering the merged buckets takes $\mathcal{O}( \bar{d} |C_{\max}|^2 )$, where $\bar{d}$ is the sum of average in- and out-degrees in $\mathcal{G}$, which is the average time required for computing the directed Jaccard for a given pair of nodes, and $C_{\max}$ is the largest cluster that is created from merging buckets.
For the discovery of emerging cliques $\mathcal{O}( \gamma |C_{\max}|^2 )$ is required in the worst case, where $\gamma$ is the fraction of nodes in the largest cluster that will not be pruned.
Overall, using $b$ bands, $\mathcal{O}\big( b [ |\mathcal{E}|r + (\bar{d} + \gamma)|C_{\max}|^2 ] \big)$  computations are required for the generation of candidate sets.
During the merging phase, %and in the absence of multiplicities, 
the complexity of processing a given candidate set is linear w.r.t. the edges adjacent to all the nodes in the set. Therefore, processing say, top $K$ candidate sets is $\mathcal{O}( \bar{d} \bar{C} K )$, where $\bar{C}$ is the average candidate set size. 
Note that the total complexity is linear in the number of edges in $\mathcal{G}$.

\noindent
{\bf Space:}  The overall space complexity  is dominated by the candidate set generation, specifically, by maintaining the similarity graph $\mathcal{G}_\mathrm{sim}$ together with its cached edges. Given $C_{\max}$, $\mathcal{O}(|C_{\max}|^2)$ memory would be required  in the worst case.

\section{Experiments}
\label{05eval}

%We evaluate \method extensively through both qualitative (Sec. \ref{ssec:qual}) and quantitative (Sec.s \ref{ssec:labeling} \& \ref{ssec:quant}) experiments.

\subsection{Setup}

\noindent{\textbf{Datasets.}} We experimented with real-world graphs of a wide variety of sizes and characteristics,
% From publicly available datasets, we used 
including
the Moreno sheep\footnote{\url{http://moreno.ss.uci.edu/data.html}} interaction network;  
a senator-to-senator network extracted from the 2009-2010 US Congress dataset \cite{conf/icwsm/Akoglu14} and the Political Blogs network \cite{adamic2005political}, both with political affiliation labels; 
DBLP 4-area\footnote{\url{https://dblp.uni-trier.de/}} co-authorship network with labels denoting author areas; 
the Cora and Citeseer  citation networks \cite{citeseer} that are labeled by publication venue; 
and finally,  transaction networks from 3 (anonymous) corporations that we collaborated with.
We also used 
unlabeled protein-protein interaction network for Homo Sapiens \cite{biogrid}; 
Stanford CS web network\footnote{\url{https://sparse.tamu.edu/Gleich/wb-cs-stanford}}; and
the Enron\footnote{\url{https://sparse.tamu.edu/LAW/enron}} email network. 
See Table \ref{tab:graphs} for a summary of network characteristics. %, where $\ast$ is used for graphs that are naturally directed but are typically treated as undirected. 

%\begin{table}[t]
\begin{wraptable}{l}{8.525cm}
	\vspace{-0.35in}
	\centering
	\caption{Real-world graphs used in experiments. $\ast$ depicts naturally directed graphs that are   typically treated as undirected. For {\small{{\tt SH, HW, KD}}} \#labels is given for EB/FS labeling.}
	\vspace{-0.05in}
	%\rowcolors{2}{}{gray!7}
	\footnotesize
	\scalebox{0.92}{
		\begin{tabular} {lrrccccc}         			
			\toprule
			
			%	Graph & \makecell{$|\mathcal{V}|$} & \makecell{$|\mathcal{E}|$} & $|\mathcal{T}|$ &  \textbf{L}abels & \textbf{D}ir.  & \textbf{M}ult. & \textbf{S}-loop   \\
			Name & \makecell{\#nodes} & \makecell{\#m-edges} & \#labels &  \textbf{L}bl. & \textbf{D}ir.  & \textbf{M}ult. & \textbf{S}-loop   \\
			\midrule 
			
			\texttt{moreno} & 0.03K & 0.25K & 2 & \textcolor{green}{\CheckmarkBold}  & \textcolor{green}{\CheckmarkBold}  & \textcolor{green}{\CheckmarkBold} &  \\
			
			%		\texttt{Macaques} & 62 & 1,187 & - & & \textcolor{green}{\CheckmarkBold}  & \textcolor{green}{\CheckmarkBold} &  \\
			
			\texttt{senate} & 0.1K & 2.4K & 2 & \textcolor{green}{\CheckmarkBold} &  &  & \\
			
			\texttt{polblogs} & 1.5K & 19K & 2 & \textcolor{green}{\CheckmarkBold}  & $^\ast$  &  & \\

			\texttt{4-area} & 27.2K & 67K & 5 & \textcolor{green}{\CheckmarkBold} &   &  & \\

			\texttt{cora} & 2.7K & 10.6K & 6 & \textcolor{green}{\CheckmarkBold} & $^\ast$ &  & \\
			
			\texttt{citeseer} & 3.3K & 9.2K & 7 & \textcolor{green}{\CheckmarkBold} & $^\ast$ &  & \\
			
			\texttt{SH trans} & 0.25K & 301K & 11/27 & \textcolor{green}{\CheckmarkBold}  & \textcolor{green}{\CheckmarkBold}  & \textcolor{green}{\CheckmarkBold} & \textcolor{green}{\CheckmarkBold}\\
			
			\texttt{HW trans} & 0.32K & 268K & 11/60 & \textcolor{green}{\CheckmarkBold}  & \textcolor{green}{\CheckmarkBold}  & \textcolor{green}{\CheckmarkBold} & \textcolor{green}{\CheckmarkBold}\\
			
			\texttt{KD trans} & 2.3K & 648K & 10/29 & \textcolor{green}{\CheckmarkBold}  & \textcolor{green}{\CheckmarkBold}  & \textcolor{green}{\CheckmarkBold} & \textcolor{green}{\CheckmarkBold}\\

			\texttt{PPI} & 3.9K & 77K & - &   &  & & \\
			\texttt{stanford} & 10K & 74K & - & & $^\ast$  & & \\		
			
			\texttt{enron} &70K & 277K & - &  & $^\ast$ & & \\
			
			\bottomrule
		\end{tabular}\label{tab:graphs}
	}\vspace{-0.4in}
\end{wraptable}

\vspace{0.1in}
\noindent{\textbf{Baselines and Configurations.}}
There is \textit{no existing method} for LMDS-graph summarization (cf. Table \ref{tab:salesman}), thus we compare only under simplified settings, w.r.t. running time and compression rate. Moreover, \method is only comparable to lossless methods.
On unlabeled simple graphs, we use {VoG} \cite{vog} as a baseline  (others in Table \ref{tab:salesman} do not have public source code available), with default settings\footnote{\url{https://github.com/GemsLab/VoG_Graph_Summarization}}. Further, we modify the \emph{Randomized} algorithm of Navlakha \etal~ \cite{navlakha} to accommodate node labels and edge directions, and compare on all graphs, ignoring the edge multiplicities. %See quantitative results in \S \ref{ssec:quant} for more details.
All experiments were run on a PC with 9th generation i-7 processor and $128$GB RAM.
\vspace{0.0in}

\subsection{Qualitative Evaluation: \method at Work}
\label{ssec:qual}

%\subsubsection{Insights through Summary Graph Visualization \& Analysis}

\noindent \textbf{Moreno sheep.} %The Moreno sheep graph 
This is a typical
complex interaction network between animals. Here, %every node is a female sheep, and 
an edge $(u, u^\prime)$ is recorded every time sheep $u$ is observed to show dominating behavior over sheep $u^\prime$. %Furthermore, the 
\begin{wrapfigure}{l}{0.59\textwidth}
	%\begin{figure}[!t]
	%\begin{tabular}{ccccc}
	\vspace{-0.1in}
	\centering
	\includegraphics[width=73mm]{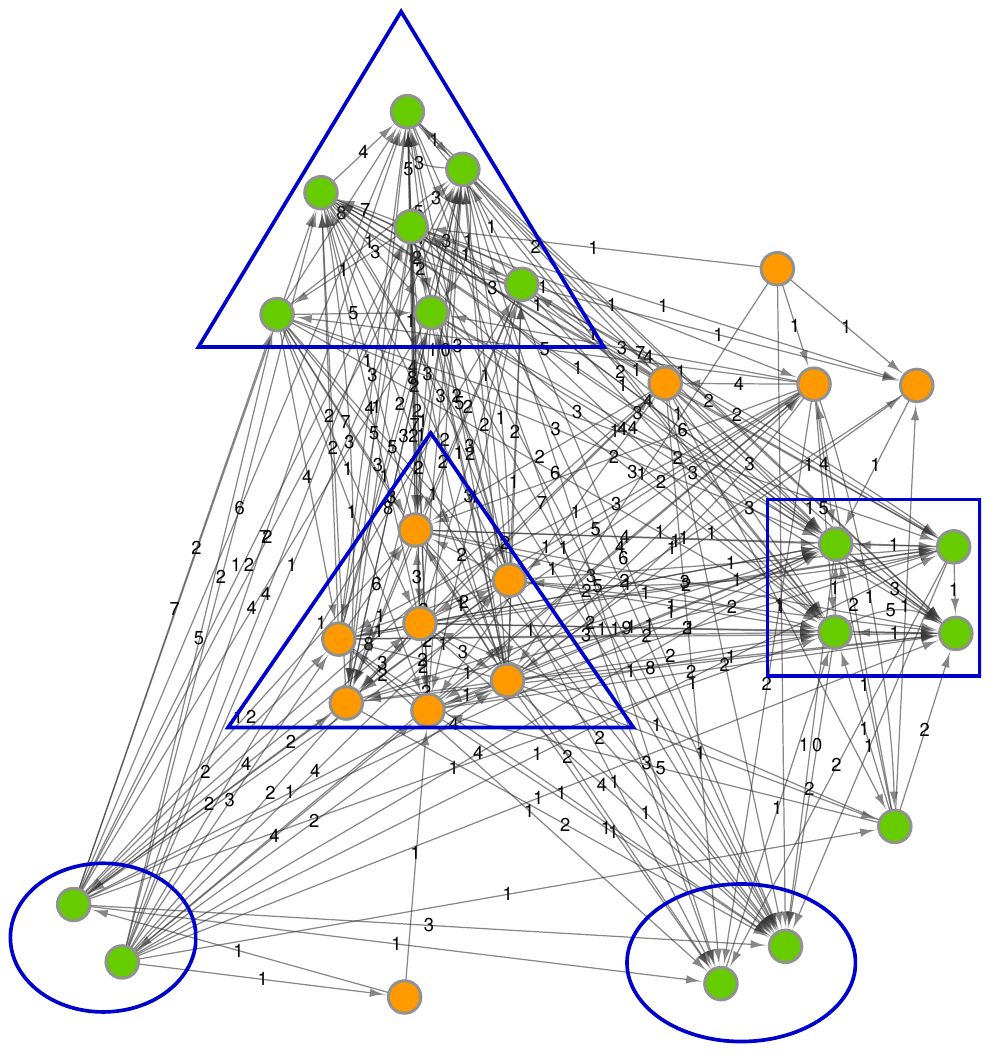}\vspace{-4cm}~
	\hspace{-3.55cm}\includegraphics[width=70mm]{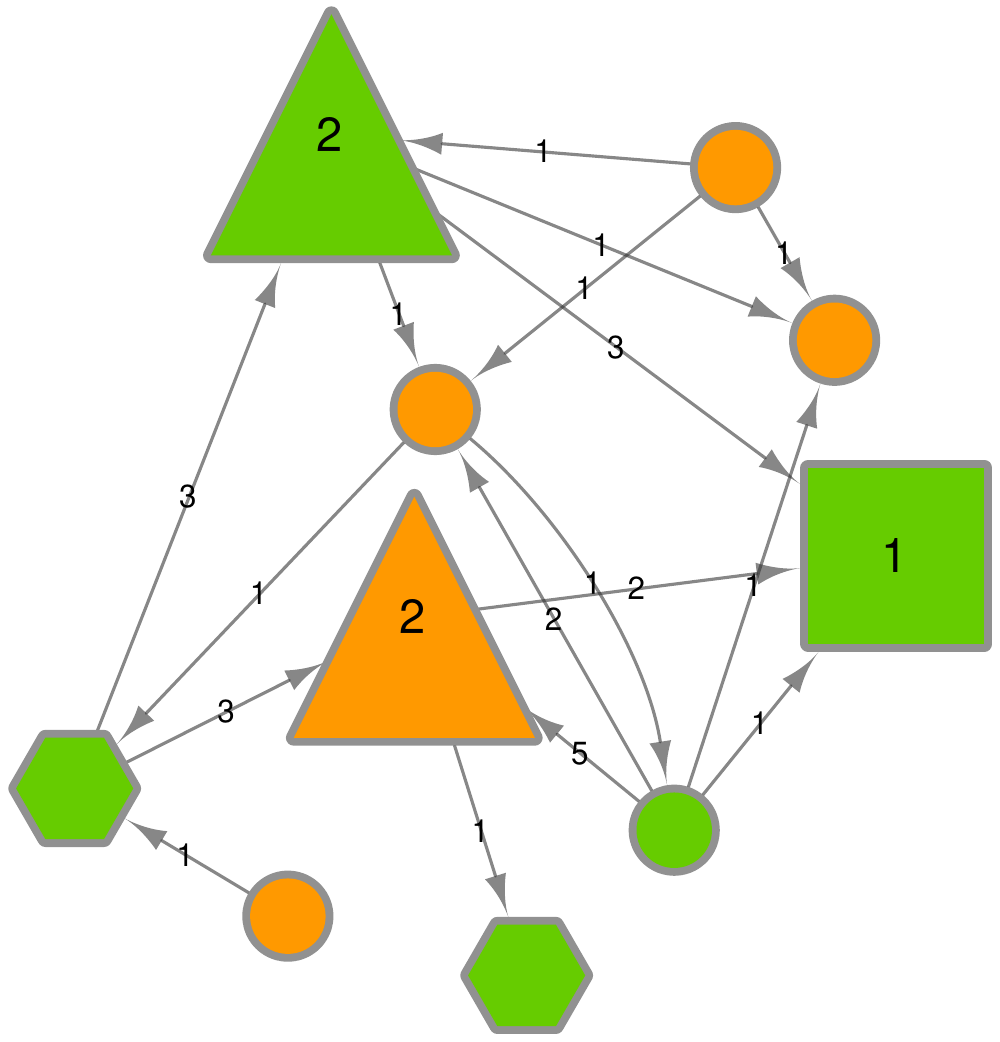}
	\vspace{-0.9in}
	\caption{Moreno sheep graph (left), summary graph (right).}
	\label{fig:sheep}
	\vspace{-0.2in}
	%\end{figure}
\end{wrapfigure}
Ages of the sheep are available, and we use them to label the individuals as young (age $\leq 7$, \textcolor{green}{green}) or old (age $>7$, \textcolor{orange}{orange}). Due to the relatively large number of recorded interactions, the full network in Fig. \ref{fig:sheep} (left) is difficult to interpret visually. The proposed \method can mitigate this problem by identifying the main structures, and building the summary graph in Fig. \ref{fig:sheep} (right).
For this type of graph, the two out-stars reveal local hierarchies for old and young sheep correspondingly, where a ``hub'' sheep dominates the spokes. The summary also includes a clique of 4 young sheep that is dominated by most other groups, and two disconnected sets, each containing a couple of sheeps that exhibit similar behavior (one couple being dominant, and the other dominated).

\begin{wraptable}{r}{0.5\textwidth}
		\vspace{-0.25in}
	\caption{ Breakdown of 4-area summary graph}
	\vspace{-0.2in}
	\begin{center}
		\small{
			\begin{tabular} {|c||c|c|c|c|c|}
				\hline
				& DB & DM & IR & ML & \textcolor{blue}{mixed}\\
				\hline
				Cliques ratio & 43$\%$ & 42$\%$ & 38$\%$ & 47$\%$ & 36$\%$ \\
				Disconnected ratio & 40$\%$ & 33$\%$ & 38$\%$ & 29$\%$ 	& 18$\%$ \\
				Single ratio & 17$\%$ & 25$\%$ & 24$\%$ & 24$\%$	& 46$\%$ \\
				%	Median Clique Size & 3 & 3 & 3 & 3 & 2 \\
				%	Max Clique Size & 26 & 13 & 	10 & 21 & 8 \\ 		
				\hline
			\end{tabular}
		}
	\end{center}
	\label{tab:coauthor}
	\vspace{-0.45in}
\end{wraptable} 

\noindent \textbf{Co-authorship.} In the 4-area coathorship network, %two authors are connected by an edge if they have co-authored a paper. Moreover, 
each author belongs to one of 4 areas (Databases, Info. Retrieval, Data Min., and Machine Learn.). We introduce a fifth category for authors that publish in more than one area. % (with various weights). 
%While the  summary graph is still too large to visualize as a whole in a meaningful way, a simple
Analysis of the resulting glyphs (see Table \ref{tab:coauthor}) reveals different connectivity patterns among the areas. For instance,  ML authors belong in cliques with higher frequency than those of other areas. 

\begin{wrapfigure}{r}{0.275\textwidth}
	\vspace{-1.2cm}
	%\centering
	
	%\begin{tabular}{ccccc}
	
	
	\hspace{1cm}\includegraphics[width=48mm]{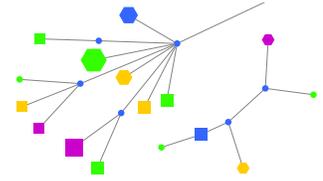}
	\vspace{-4.4cm}
	\caption{4-area coauthorship graph summary visualized partly. Mixed-area authors (\textcolor{blue}{blue}) tend to ``bridge'' authors from a single area. (best in color)}
	\vspace{-0.7cm}
	\label{fig:4area}
\end{wrapfigure} 
On the other hand, DB authors have a higher tendency to form disconnected groups that collaborate with one author. 
Of high interest is the case of the fifth category of interdisciplinary (mixed-area) authors, %. Authors in this category 
whom remain single (unmerged) at a significantly higher ratio than single-area authors. Visual inspection of different parts of the summary (see e.g. Fig. \ref{fig:4area}) reveals that the reason for mixed-area authors not being merged is because they frequently act as ``bridges'' that connect authors from two or more different areas. This showcases the important role of interdisciplinary researchers as the ``glue'' of the academic community.

\begin{figure}[!t]
	%\begin{tabular}{ccccc}
	\vspace{-0.1in}
	\centering
	\includegraphics[width=50mm]{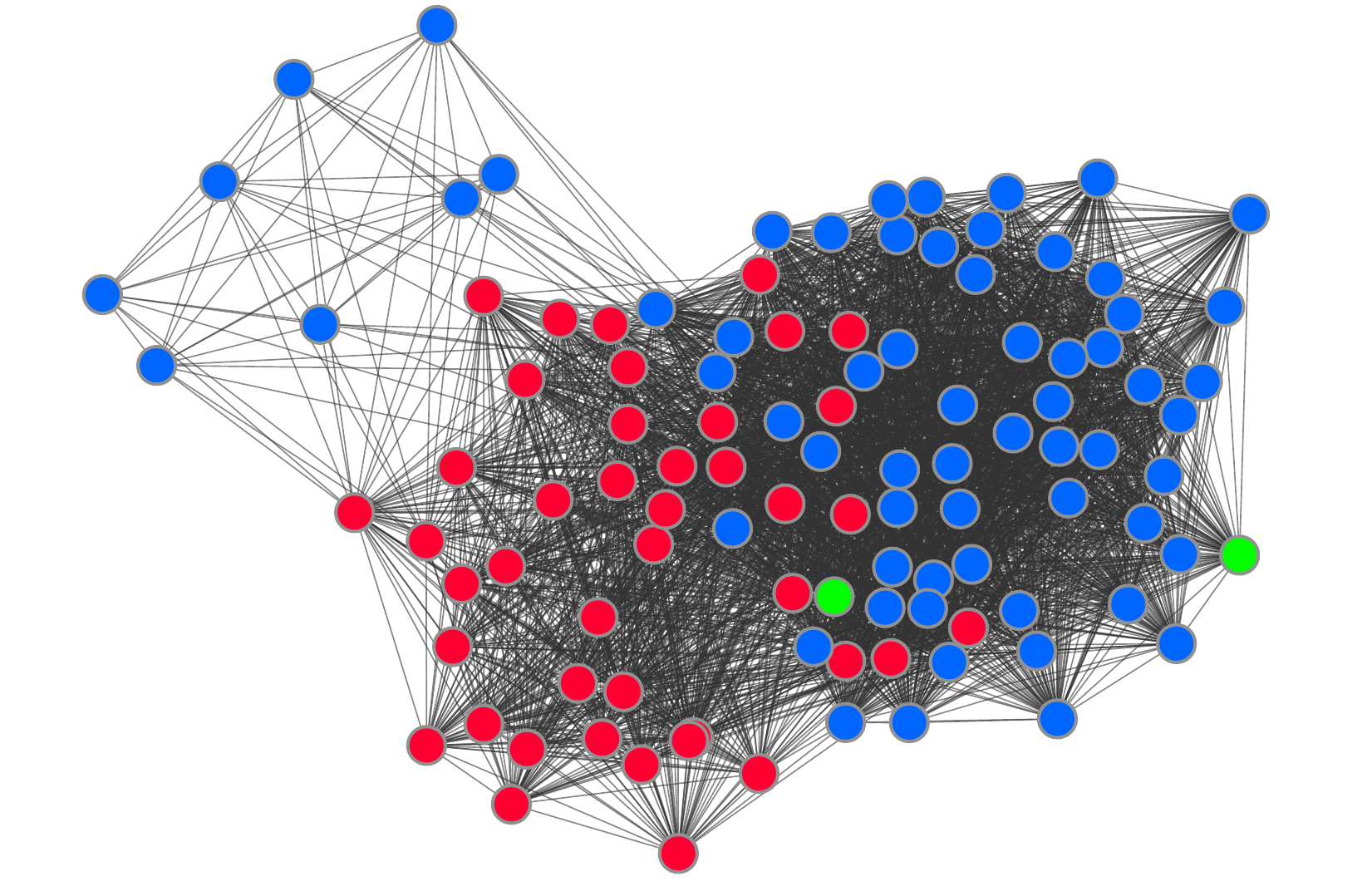}
	\includegraphics[width=40mm]{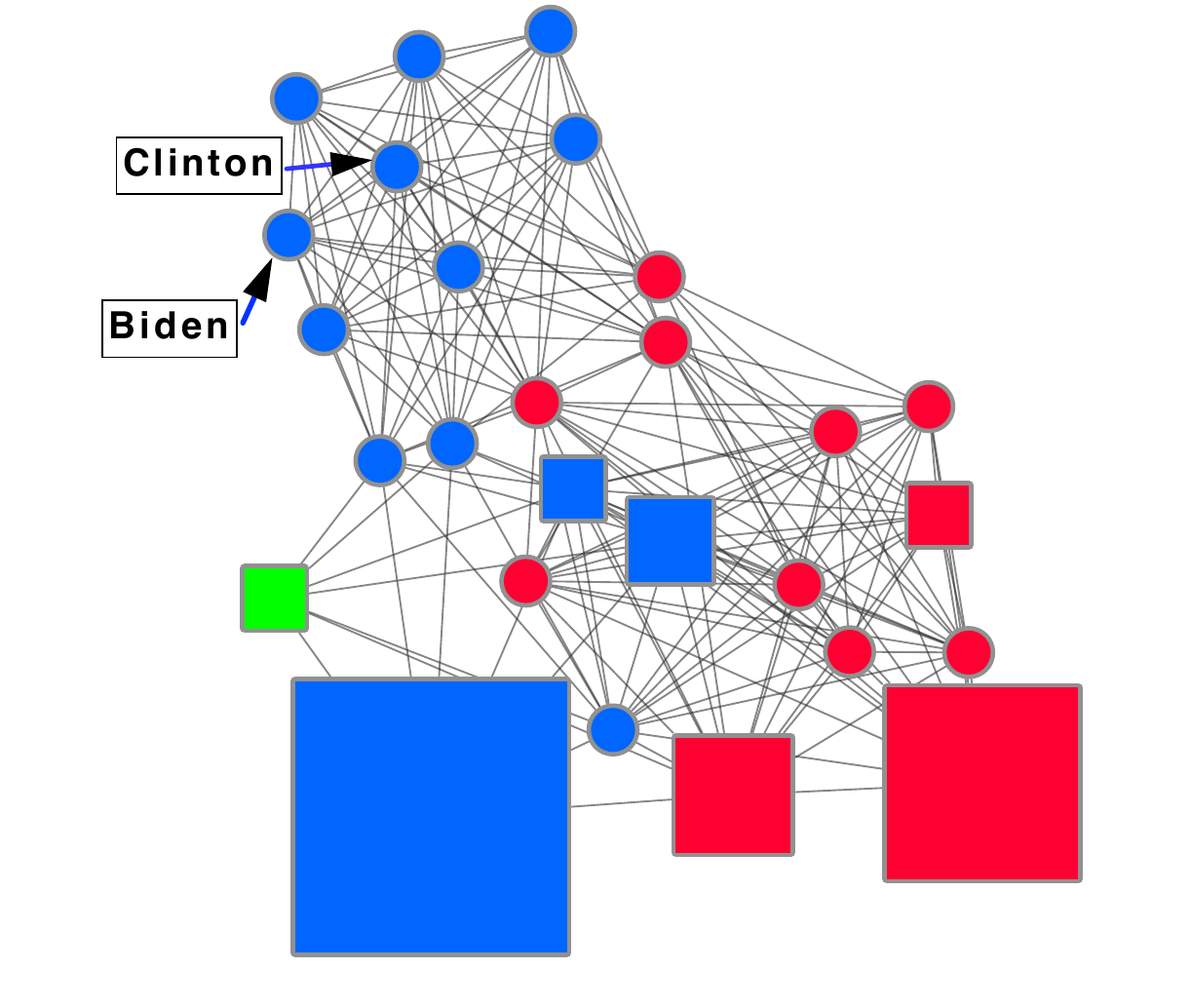}~
	\includegraphics[width=50mm]{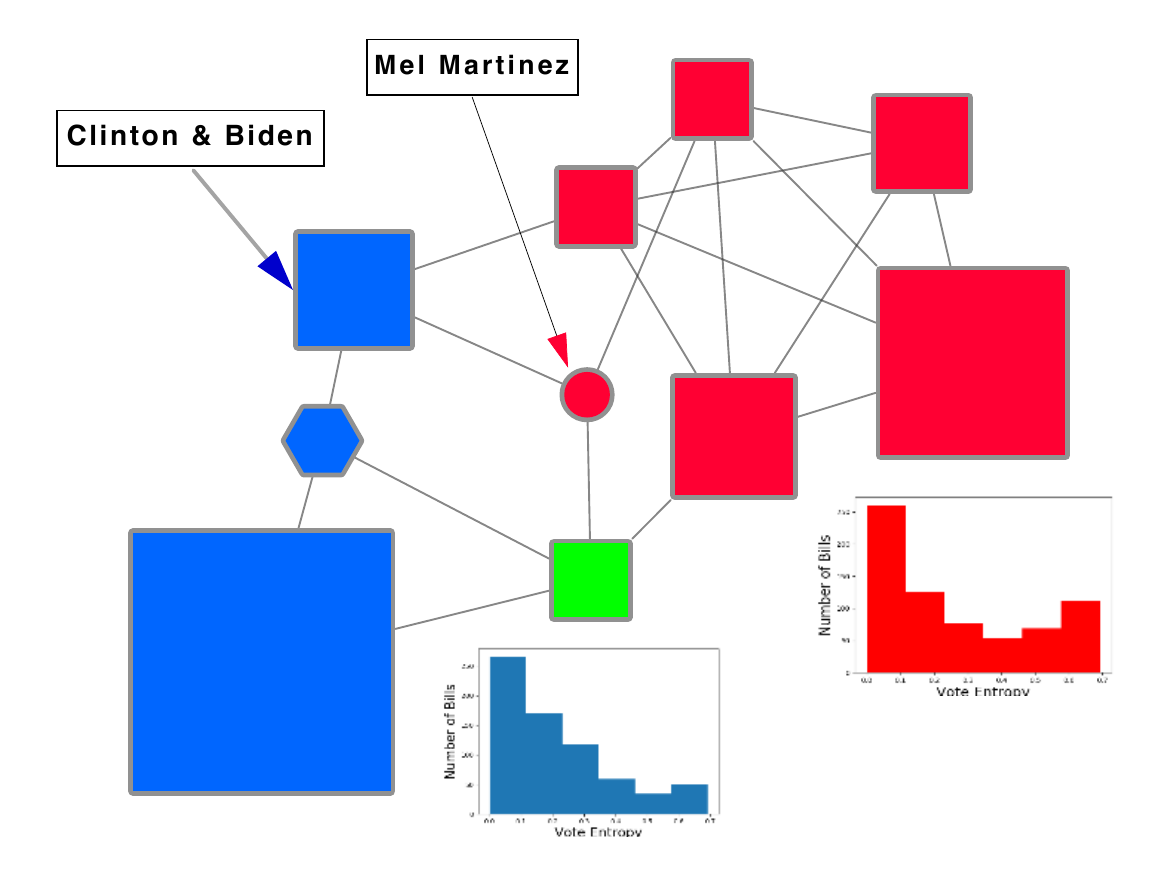}
	\vspace{-0.05in}
	\caption{(left) original US Senate graph, (middle) high resolution ($b=2$) summary, (right) low resolution ($b=5$) summary.}
	\label{fig:senate}
	\vspace{-0.15in}
\end{figure}

\noindent \textbf{Senate.} %Using the US Senate dataset which 
The dataset contains the (positive or negative) votes of 108 senators for 696 congressional bills. The senators are labeled as Republicans (\textcolor{red}{red}), Democrats (\textcolor{blue}{blue}), or independent (\textcolor{green}{green}). We construct an undirected graph where two senators are connected by an edge if the cosine similarity of their votes is larger than $0.3$. The graph is plotted in Fig. \ref{fig:senate}, along with two summaries at different resolutions, leading to the following observations. 
Interestingly, while most democratic senators eventually form a clique, there is a smaller group of East coast senators, including prominent Democratic figures such as Joe Biden, Hillary Clinton, and Ted Kennedy that do not merge with the main body and form their own separate clique. Furthermore, this clique of Democrats is directly linked to certain Republicans, such as the Florida-based Mel Martinez, who has most recently opposed Trump openly and explicitly expressed his preference for Joe Biden\footnote{\small\url{https://theweek.com/speedreads/609406/former-rnc-chairman-says-wont-vote-trump-wishes-joe-biden-run}}.
The second observation is that Republican senators overall exhibit a more fragmented voting behavior, splitting into multiple cliques of comparable size. This is corroborated by computing the entropies of the votes for all the bills, for Democrats and Republicans separately. Intuitively, bills with high entropy indicate a low degree of agreement on the subject. By plotting the histograms of the voting entropies (see Fig. \ref{fig:senate} (right)) for the two groups, it becomes apparent that Republican votes exhibit higher entropy (median = $0.21$) than Democrats (median = $0.16$). Moreover, Republicans have more than $100$ highly-contested bills (entropy$\approx 0.7$), while the same for Democrats is less than $50$. 

\subsection{Quantitative Evaluation: Evaluating Financial Accounts Labeling}
\label{ssec:labeling}
In this section, we show how we employed \method to quantitatively address a domain-specific problem, specifically, evaluating a pre-existing \textit{labeling}, i.e., the set of types pre-assigned to the nodes in an %graph. One such application is in business accounting, 
%where a 
accounting network that connects business accounts via credit/debit transaction relations.% the double-entry bookkeeping system.

A business entity's Chart of Accounts (COA) lists, and also pre-assigns a label to, each distinct account used in its ledgers. Such labeling helps companies prepare their aggregate financial statements (FS). 
For example, the FS caption ``Cash and Cash Equivalents'' is used to describe the total sum of all liquid assets tracked in a number of accounts; e.g., currencies, checking accounts, etc.
In the US, FS captions are not uniform across corporations. In fact, the data we have from 3 different companies (anonymized as SH, HW, and KD in Table \ref{tab:graphs}) each contains different FS captions. 

How suitable is a given FS labeling?\footnote{Evaluating COA is beneficial, as it can become stale when corporations change due to market conditions or acquisitions and mergers. It can also assess any proposed new COA labeling on past data before implementation of the new system.} 
Can a different labeling be shown to be quantitatively better than another?

To this end, our collaborator (an accounting expert) designed a new labeling (referred as EB for economic bookkeeping), relabeling the accounts based on their primary economic nature. Specifically, EB organizes them into operating versus financing and long- versus short-term accounts.
Expert knowledge suggests that EB improves over FS captions in a couple of ways. First, EB splits the accounts under a generic FS caption like ``inter-company accounts'' to reflect the financial or operating nature of within-firm activities. EB also consolidates excessive ``over-labeling'' in FS, combining a few economically similar FS captions into a single label; e.g., combining ``accounts payable'' and ``accrued expenses'' into ``short-term operating liabilities''.

\noindent \textbf{Transaction networks.}
Ideally, accounts of the same label should ``behave'' similarly in the system. This behavior can be discerned from the real-world usage data, in particular the transactions graph, where accounts are connected through credit/debit relations.  Under a more suitable labeling, the accounts with the same label should have more structural similarity and yield better compression.

%\begin{wraptable}[!t]
\begin{wraptable}{l}{6.25cm}
		\vspace{-0.7cm}
	\caption{Evaluating account labelings in financial networks}
	\vspace{-0.25in}
	\begin{center}
		\scriptsize{
			\begin{tabular} {|c|c||c|c|c|}
				\hline
				Dataset	& Labeling & Shuffled & Actual  & norm. gain ($\%$) \\
				\hline
				\multirow{2}{*}{SH} 
				& EB   & 0.28 & 0.32 & \textbf{5.6} $\%$ \\
				& FS   & 0.25 & 0.27 & 2.7 $\%$ \\
				\hline
				\multirow{2}{*}{HW} 
				& EB   & 0.36 & 0.47  & \textbf{17.0} $\%$ \\
				& FS   & 0.16 & 0.27 & 13.0 $\%$ \\
				\hline
				\multirow{2}{*}{KD}
				& EB   & 0.33 & 0.42 & \textbf{13.7} $\%$ \\
				& FS   & 0.31 & 0.39 & 12.0 $\%$ \\
				\hline
			\end{tabular}
		}
	\end{center}
	\label{tab:transaction}
	\vspace{-0.4in}
\end{wraptable}
To compare EB vs. FS, we employ \method on each graph using one or the other labeling separately, and record the compression rate. % (see Eq. \eqref{rate} below).
 Next, we shuffle the labels (within each setting) randomly, and employ \method again. ``Shuffled'' and ``Actual''  compression rates are reported in Table \ref{tab:transaction} (the former averaged over 20 random shuffles). EB rates are higher---this is not surprising as EB has fewer labels as compared to FS (See $|\mathcal{T}|$ in Table \ref{tab:graphs}), and hence \method has higher degree of freedom to merge nodes on EB-labeled graphs. As such, Actual values are not directly comparable. What is comparable is the difference from Shuffled, that is, how much the labeling can improve on top of the random assignment of the \textit{same} set of labels. Here, the absolute difference is always equal or larger for EB. 

However, even the absolute difference of compression rates is not fair to compare---it is harder to compress a graph that has been compressed quite a bit even further. For EB, Shuffled rates are already high. Improving over Shuffled even by the same amount proves EB superior to FS.  
Therefore, we report the normalized gain; defined as (Actual$-$Shuffled) / (1$-$Shuffled), which shows that our expert-designed EB labeling is better, for the aforementioned reasons. 
All in all, \method can be employed as a data-driven tool to quantitatively assess the connection between a labeling and the actual role of nodes in practice as reflected in the data.

\subsection{Quantitative Evaluation: Compression rate, Running time, Scalability}
\label{ssec:quant}

%A summary $\mathcal{G}_\mathrm{s}$ of a given graph $\mathcal{G}$ has total encoding cost $\mathrm{Total~Bits}= \mathrm{Bits}(\mathcal{G}_\mathrm{s}) + \mathrm{Bits}(\mathcal{G}|\mathcal{G}_\mathrm{s})$ which is measured in bits.
Quantitatively, we measure summarization performance in terms of both (1) running time, as well as (2) the size reduction achieved in terms of bits (including bits required for correction).
Specifically, upon obtaining the total number of bits (as given by the encoding scheme of each method), we measure the compression ratio as
%\small
%\vspace{-0.15in}
%\begin{equation}
%\label{rate}
%\hspace{-0.03in}
$
\mathrm{Compress~Ratio} = \frac{ \mathrm{Bits\_Before} - \mathrm{Bits\_After} }{\mathrm{Bits\_Before} } \in [0,1)
$, 
%\end{equation}
%\normalsize
that is the fraction of the encoding cost %of the original graph 
that has been reduced by summarization. We compare with the Navlakha algorithm \cite{navlakha}, which we modified to handle edge directions and node labels. %, which produces a summary as well as corrections. % with cost that can be quantified using our encoding. 
We also compare against VoG only on unlabeled simple (i.e., unweighted)  graphs (see Table \ref{tab:salesman}), where cost is also measured in bits \cite{vog}. %For TG-sum, we set $r=2$ for all graphs, except for Polblogs ($r=1$) and Enron ($r=5$), and keeping the $K=5,000$ first candidate sets, except for Enron where we kept $k=15,000$. 
We run \method by gradually increasing $b$, to increase the number of candidate sets and obtain multi-resolution summaries.  A larger number of candidates is expected to yield higher compression ratio, albeit at the cost of increased running time---hence enabling the user to choose a suitable trade-off in practice.

\begin{figure}
	\vspace{-0.2in}
	\centering
	
	\begin{tabular}{cccc}
		
		
		\includegraphics[width=35mm]{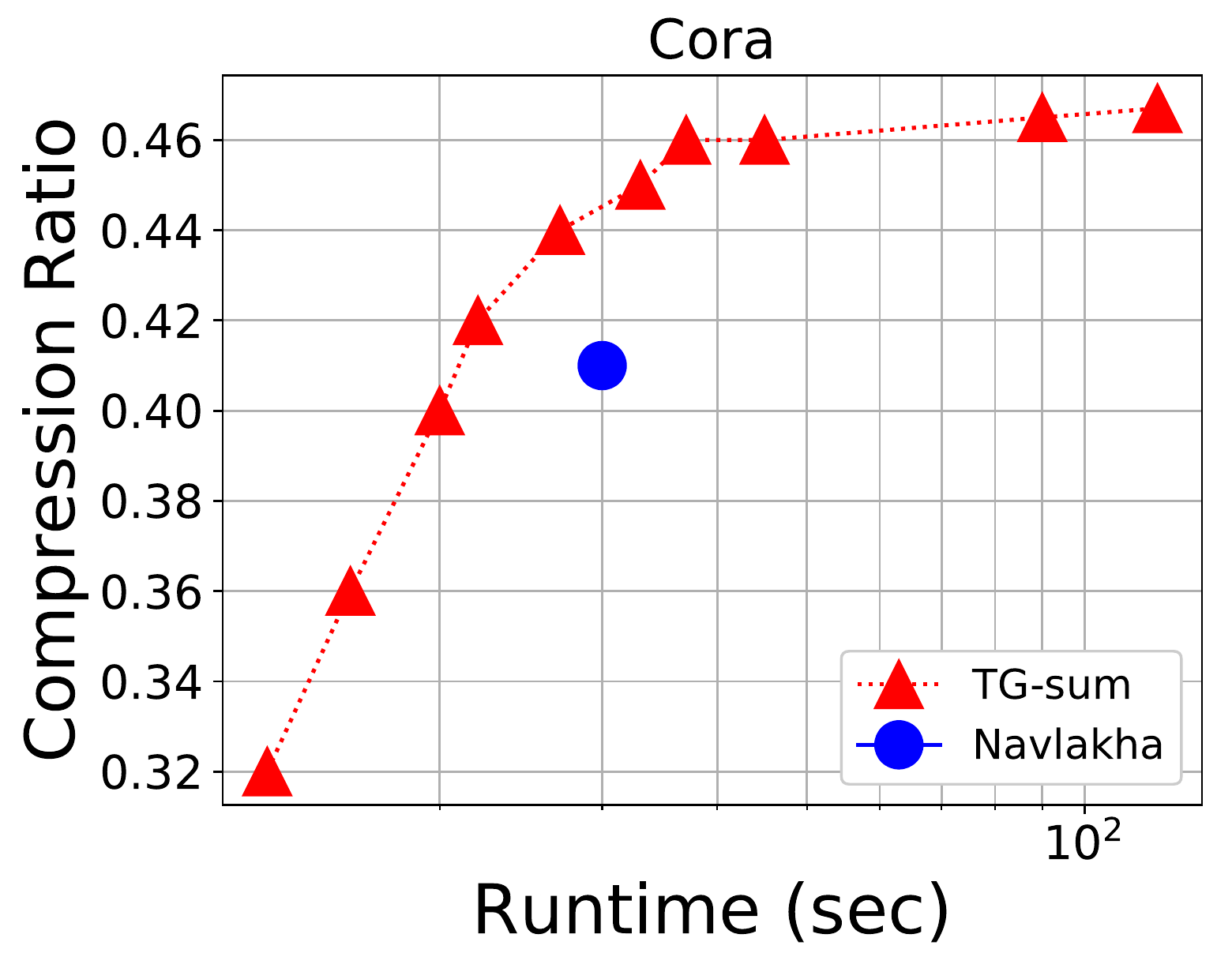}~ &
		\includegraphics[width=35mm]{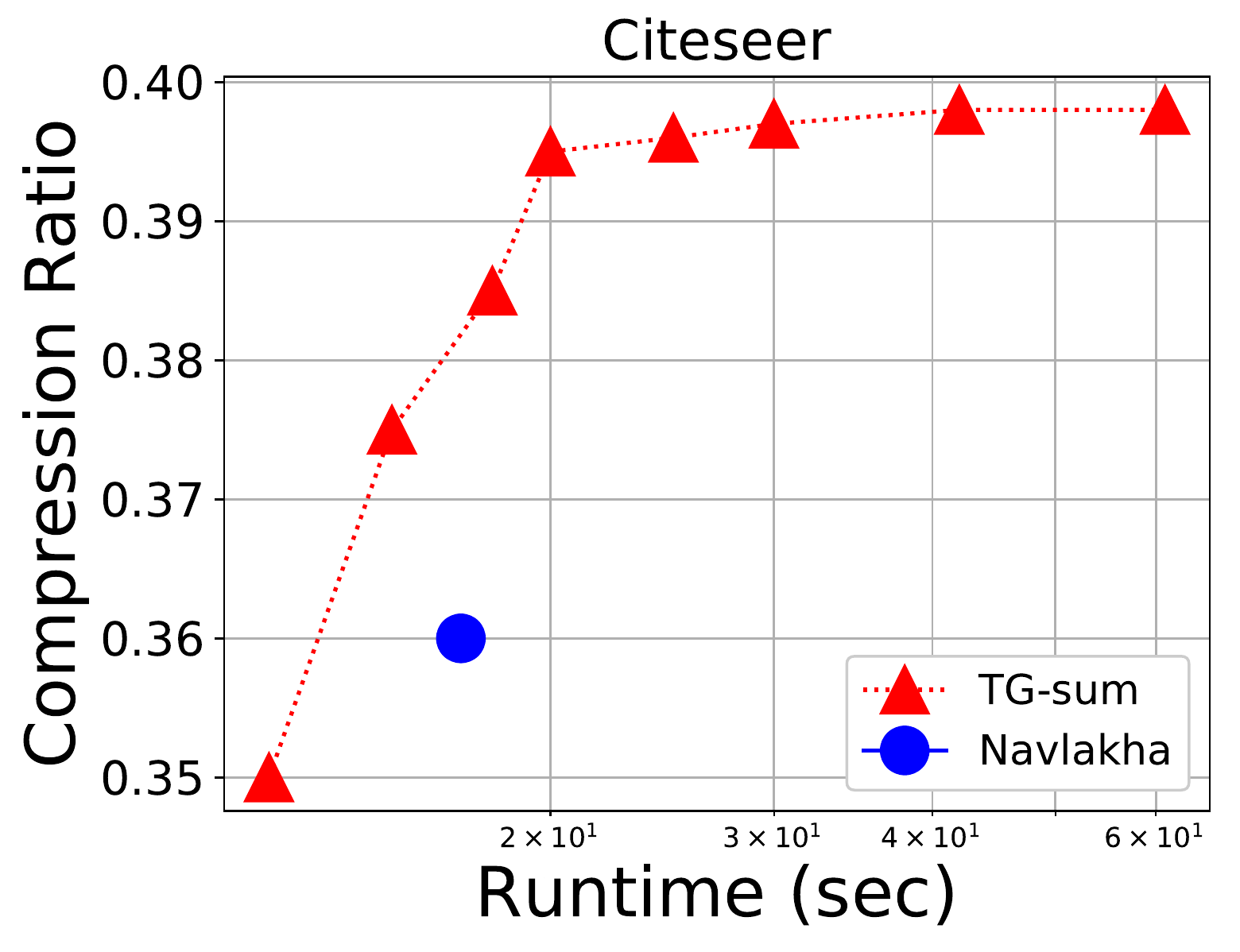} &
		\includegraphics[width=35mm]{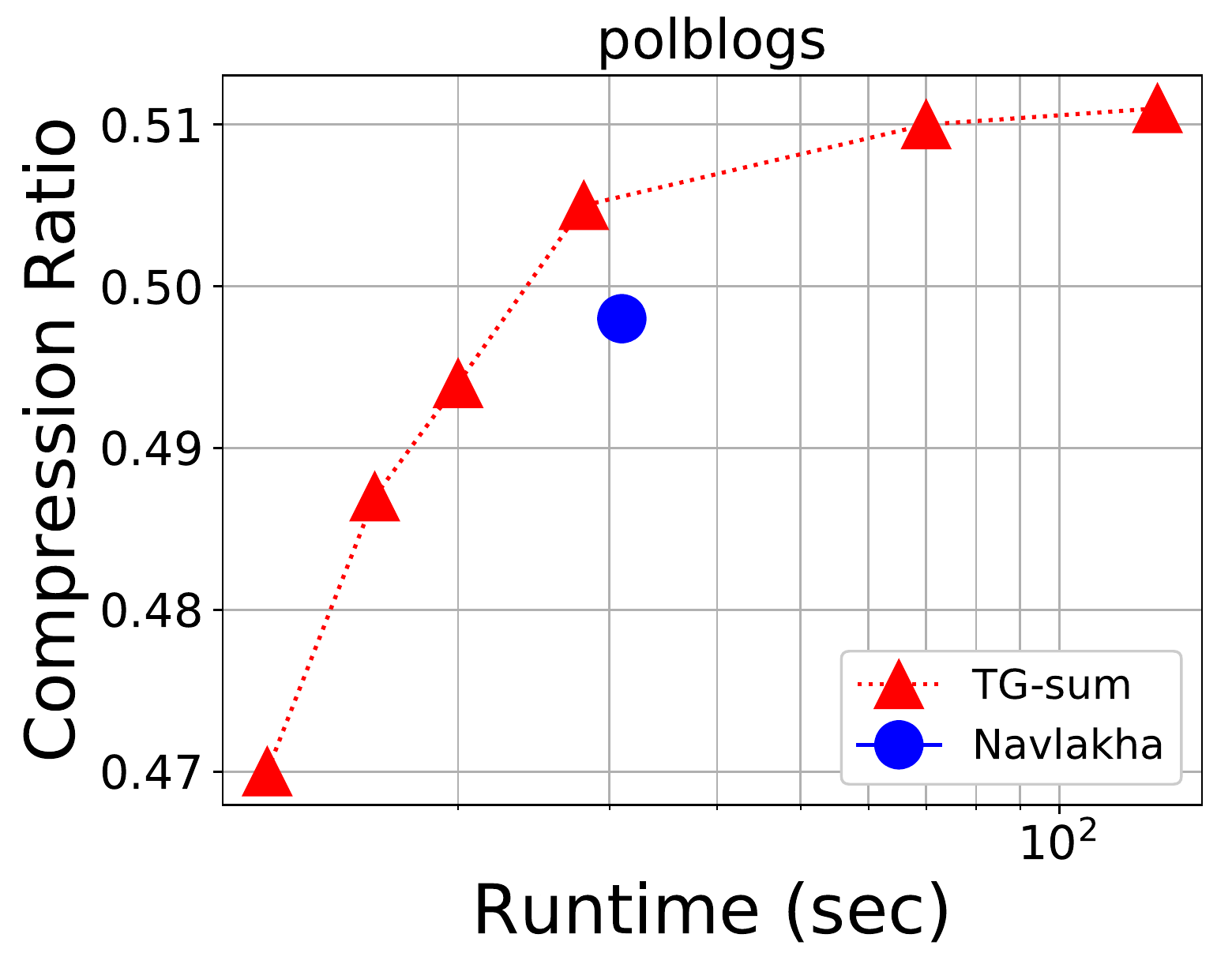} &
		\includegraphics[width=35mm]{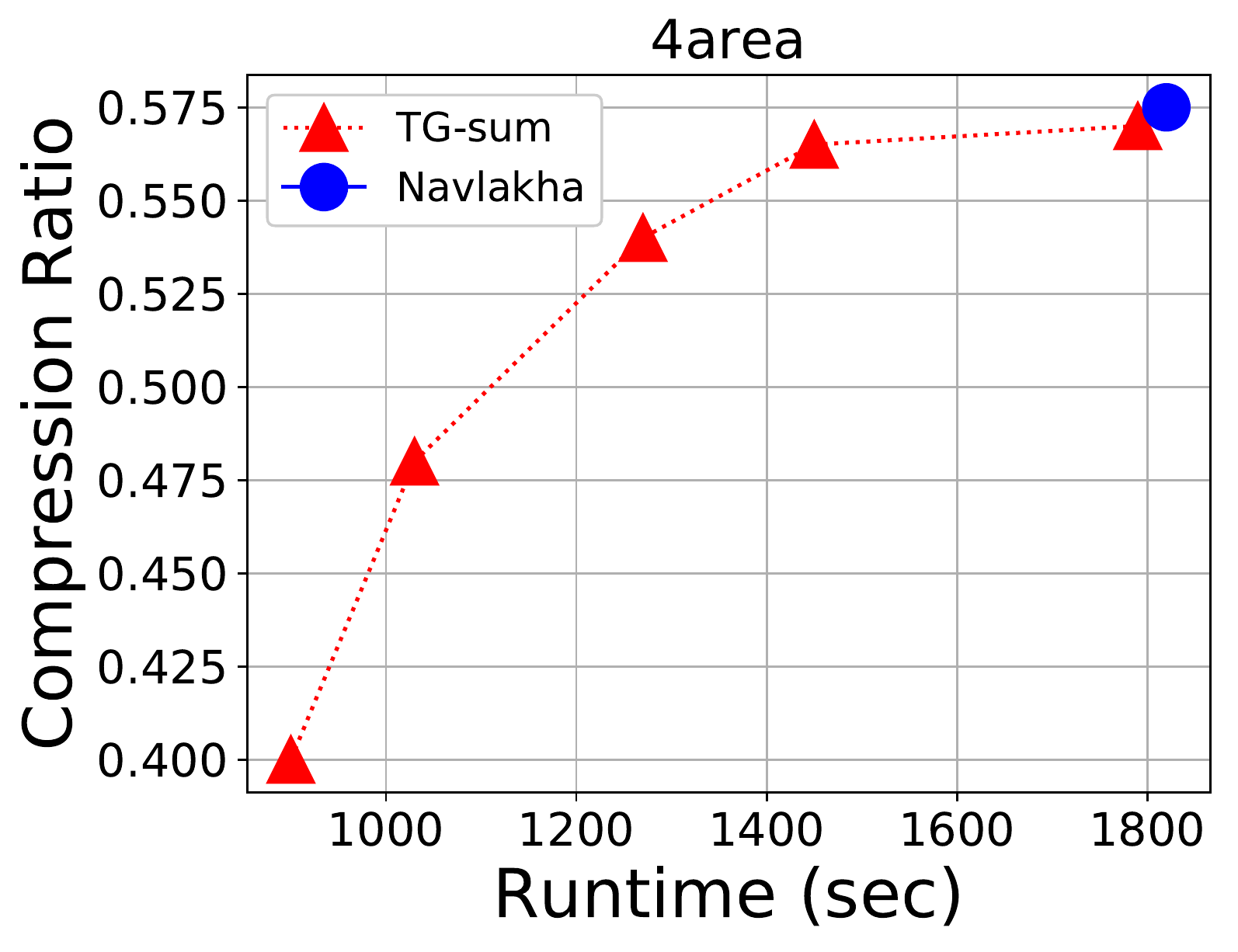}
	\end{tabular}
	\caption{Compression ratio vs. runtime on \underline{un}directed graphs \label{fig:labeled}}
	\vspace{-0.1in}
\end{figure}

%\vspace{-1.0in}

\begin{wrapfigure}{l}{0.59\textwidth}
%\begin{wrapfigure}[t]
\begin{tabular}{ccc}
	\hspace{-0.1cm}
	\includegraphics[width=28mm]{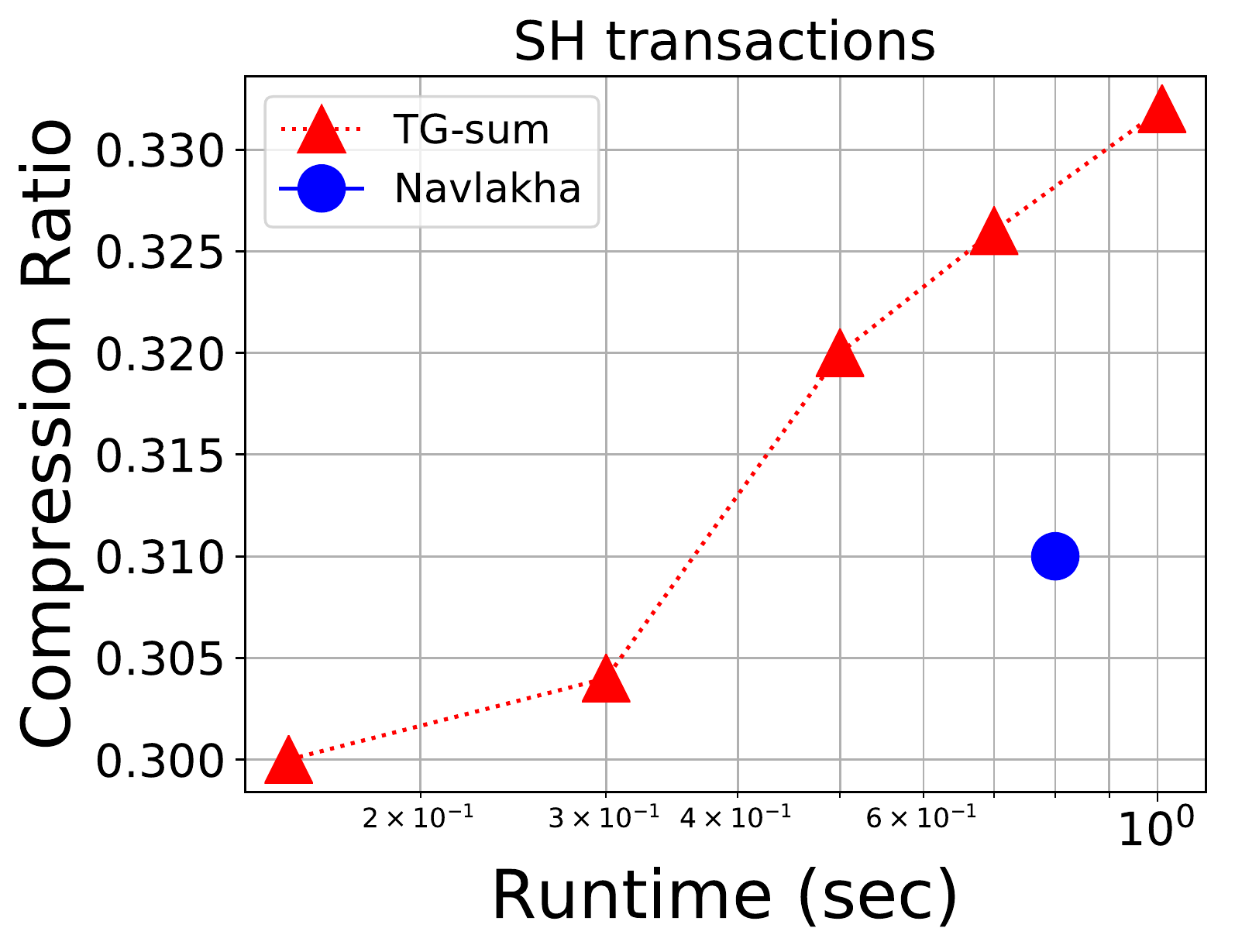}
	&
	\hspace{-0.05cm}\includegraphics[width=28.5mm]{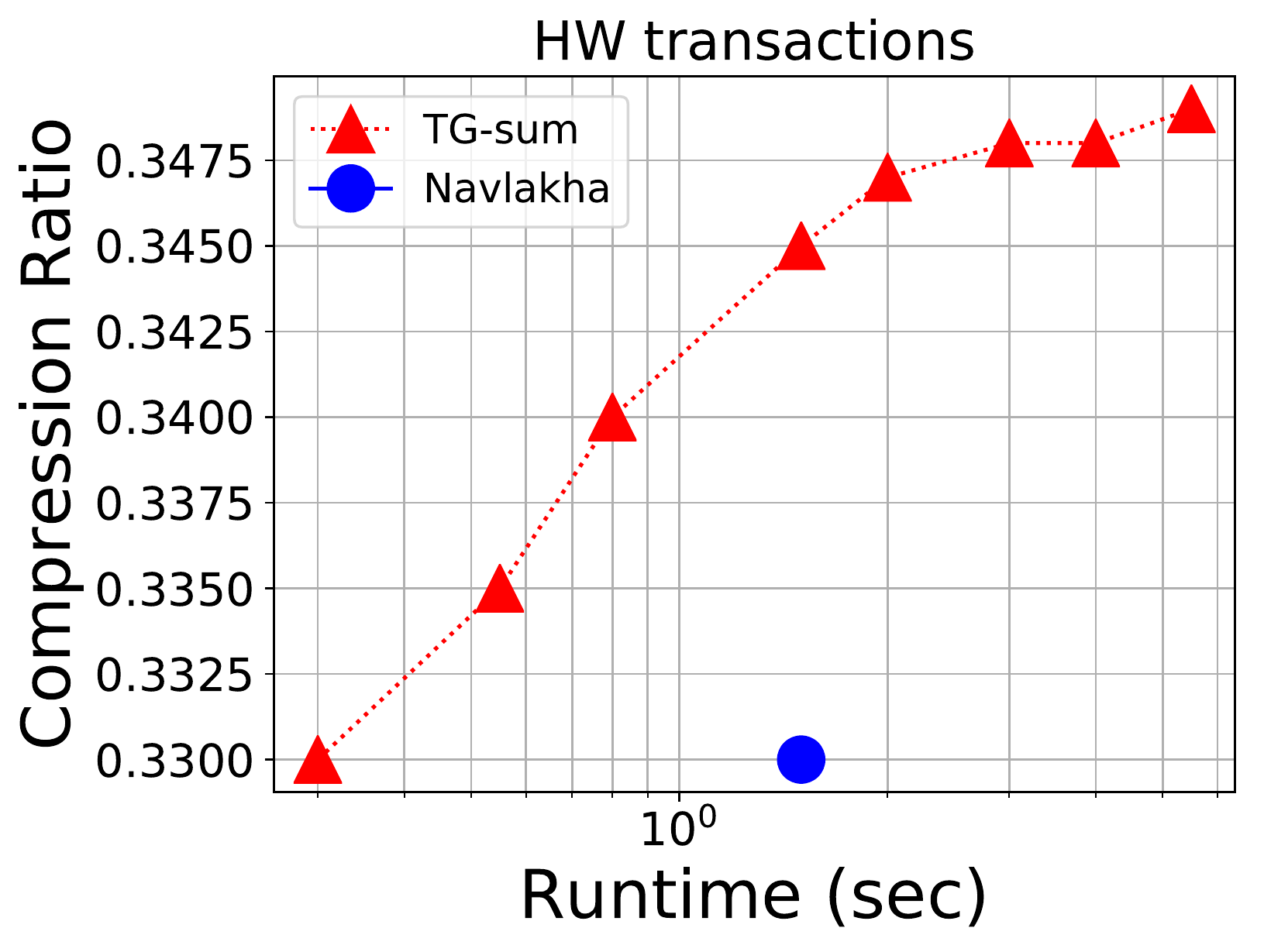}
	&
	\hspace{-0.05cm}\includegraphics[width=28mm]{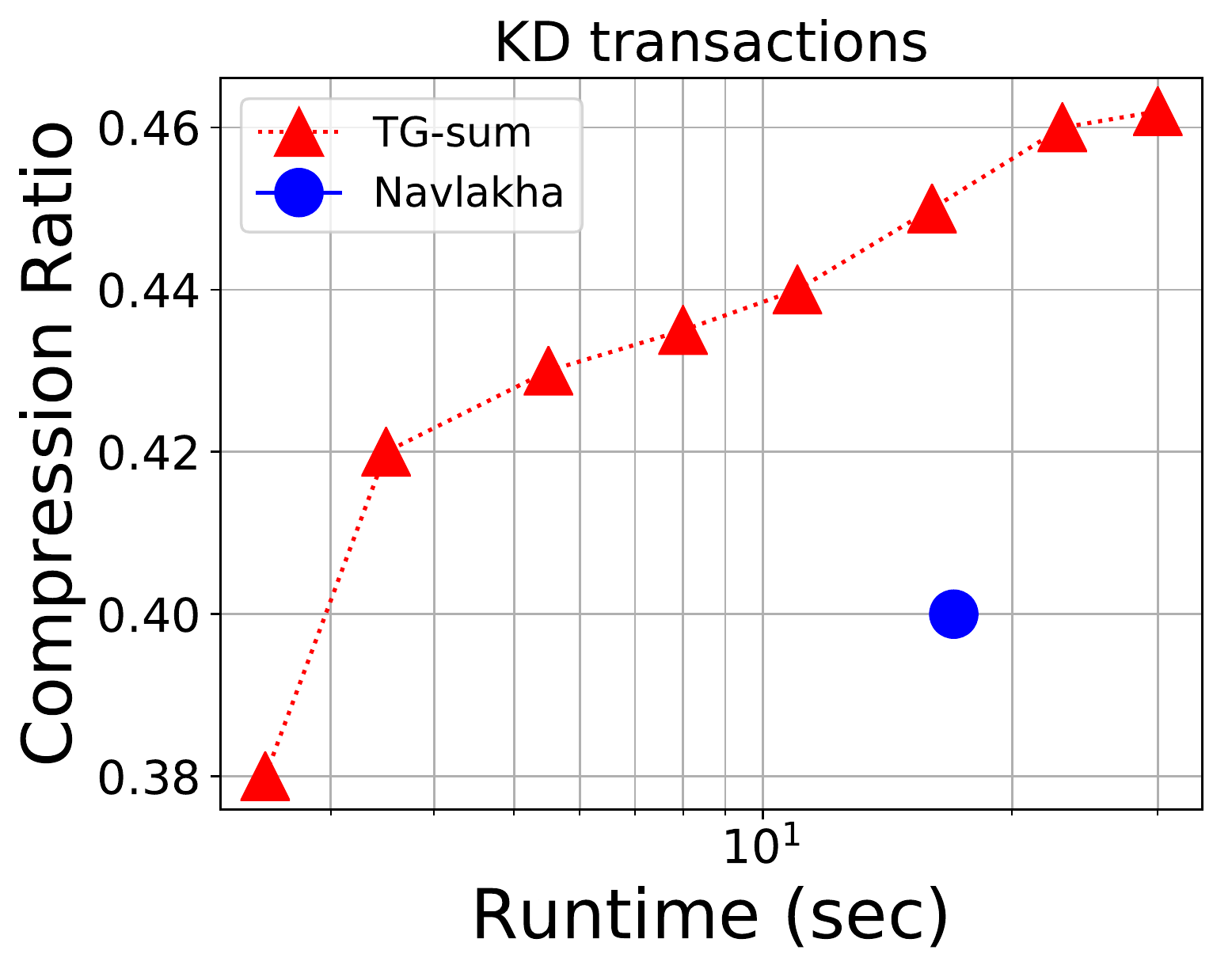}\\
	
	\multicolumn{3}{c}{(a) LMDS graphs} \\
	
		\hspace{-0.1cm}\includegraphics[width=28mm]{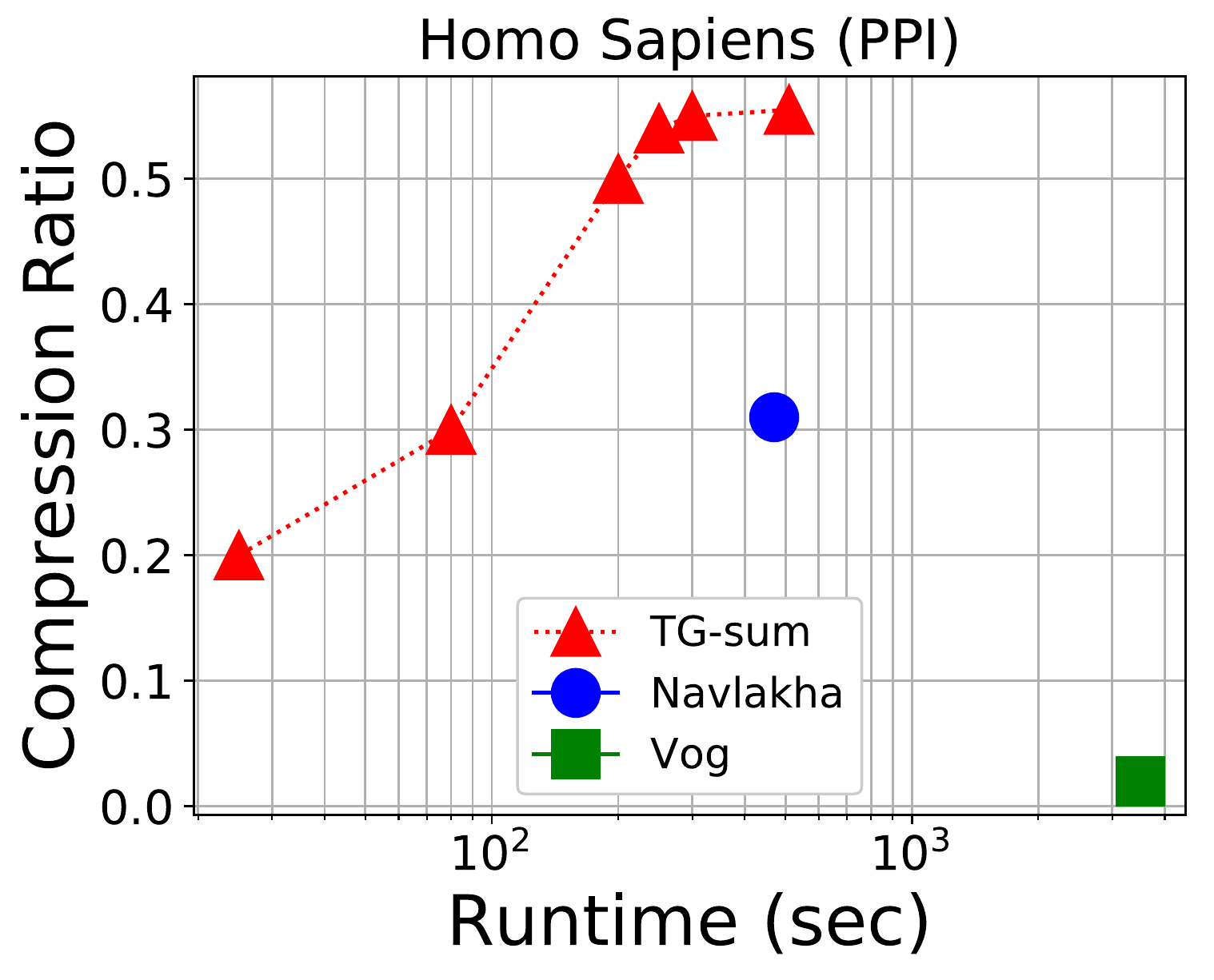}
		&
	\hspace{-0.05cm}\includegraphics[width=28mm]{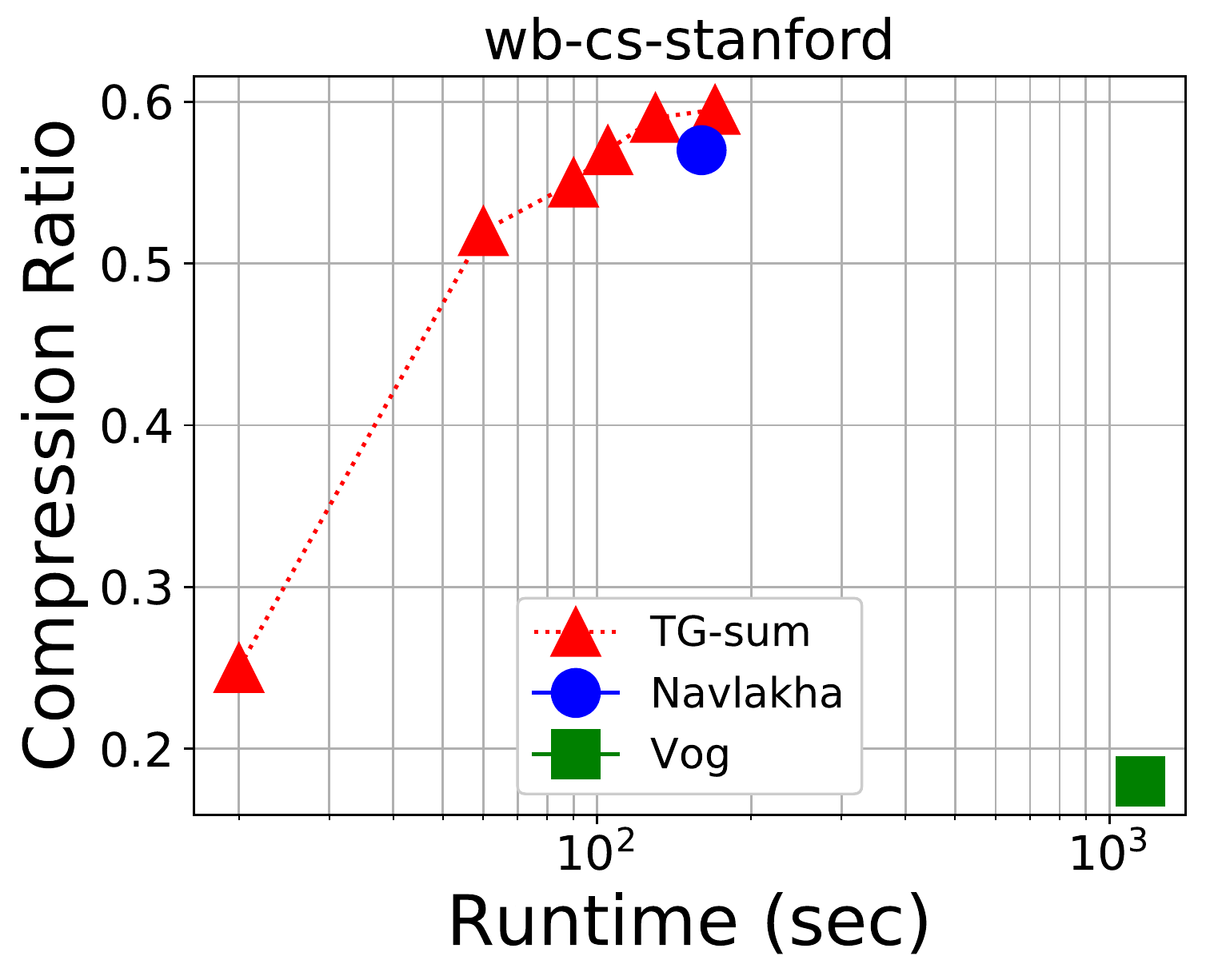}
	&
	\hspace{-0.05cm}\includegraphics[width=29mm]{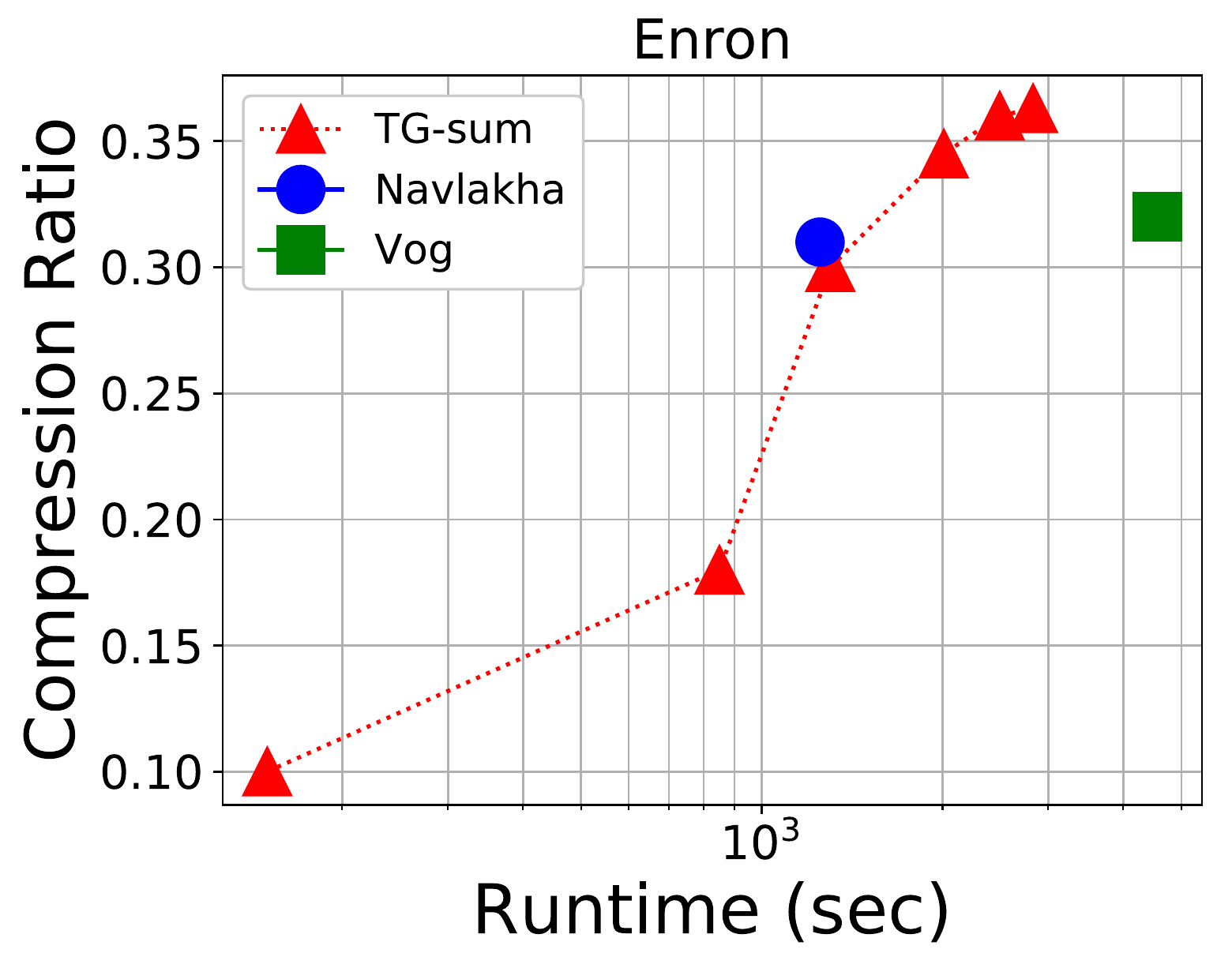} \\
	
	\multicolumn{3}{c}{(b) \underline{un}labeled \underline{simple} graphs}
	
\end{tabular}
	\vspace{-0.05in}

	\caption{Compression ratio vs. runtime 	\label{fig:trans}}
	\vspace{-0.2in}
\end{wrapfigure}

%\vspace{-1.0in}
Results are given in Fig. \ref{fig:labeled} (for labeled \textit{undirected} graphs), in Fig. \ref{fig:trans}(a) (for LMDS graphs),  and in Fig. \ref{fig:trans}(b) (for  \textit{unlabeled} \textit{simple} graphs).
\method remarkably outperforms the alternatives in terms of compression ratio in almost all cases. In other cases, it achieves comparable compression within similar or even lower running time. 
In absolute terms, it achieves roughly 30--60\% compression across these various real-world graphs with up to hundreds of thousands of edges and tens of distinct labels.
%The compression improvement  saturates after some point, due to no new candidates being generated, or them being of poor quality and rejected by the subsequent candidate selection.

%\begin{wrapfigure}{l}{0.59\textwidth}
%%\begin{wrapfigure}[!h]
%	\vspace{-0.2in}
%	%\centering
%	
%	%\begin{tabular}{ccccc}
%	
%	% Requires \usepackage{graphicx}
%	
%	%	\includegraphics[width=38mm]{figs/quant/sh.eps}~\hspace{-0.3cm}\includegraphics[width=38mm]{figs/quant/homo.eps}\\
%	%		\hspace{-0.3cm}\includegraphics[width=38mm]{figs/quant/hw.eps}~\hspace{-0.3cm}\includegraphics[width=38mm]{figs/quant/stanford.eps}\\
%	%    \hspace{-0.3cm}\includegraphics[width=38mm]{figs/quant/kd.eps}~\hspace{-0.3cm}\includegraphics[width=38mm]{figs/quant/enron.eps}
%	\hspace{-0.1cm}\includegraphics[width=28mm]{figs/quant/homo.eps}~
%	\hspace{-0.05cm}\includegraphics[width=28mm]{figs/quant/stanford.eps}
%	\hspace{-0.05cm}\includegraphics[width=29mm]{figs/quant/enron.eps}
%	%\end{tabular}
%	\vspace{-0.05in}
%	\caption{Compr. ratio vs. runtime on \underline{un}labeled \underline{simple} graphs \label{fig:simple}}
%	%\vspace{-0.4in}
%\end{wrapfigure}

Finally, we measure the scalability of \method by first generating an increasing size synthetic directed $k$-out graph, where nodes are incrementally added and connected to $k=10$ of the existing nodes, simulating a preferential attachment process \cite{jeong2003measuring}.  

\begin{wrapfigure}{r}{0.6\textwidth}
	%\begin{wrapfigure}[b]
	\vspace{-0.35in}
	\centering
	
	%\begin{tabular}{ccccc}
	
	\hspace{-0.23cm}\includegraphics[width=47mm]{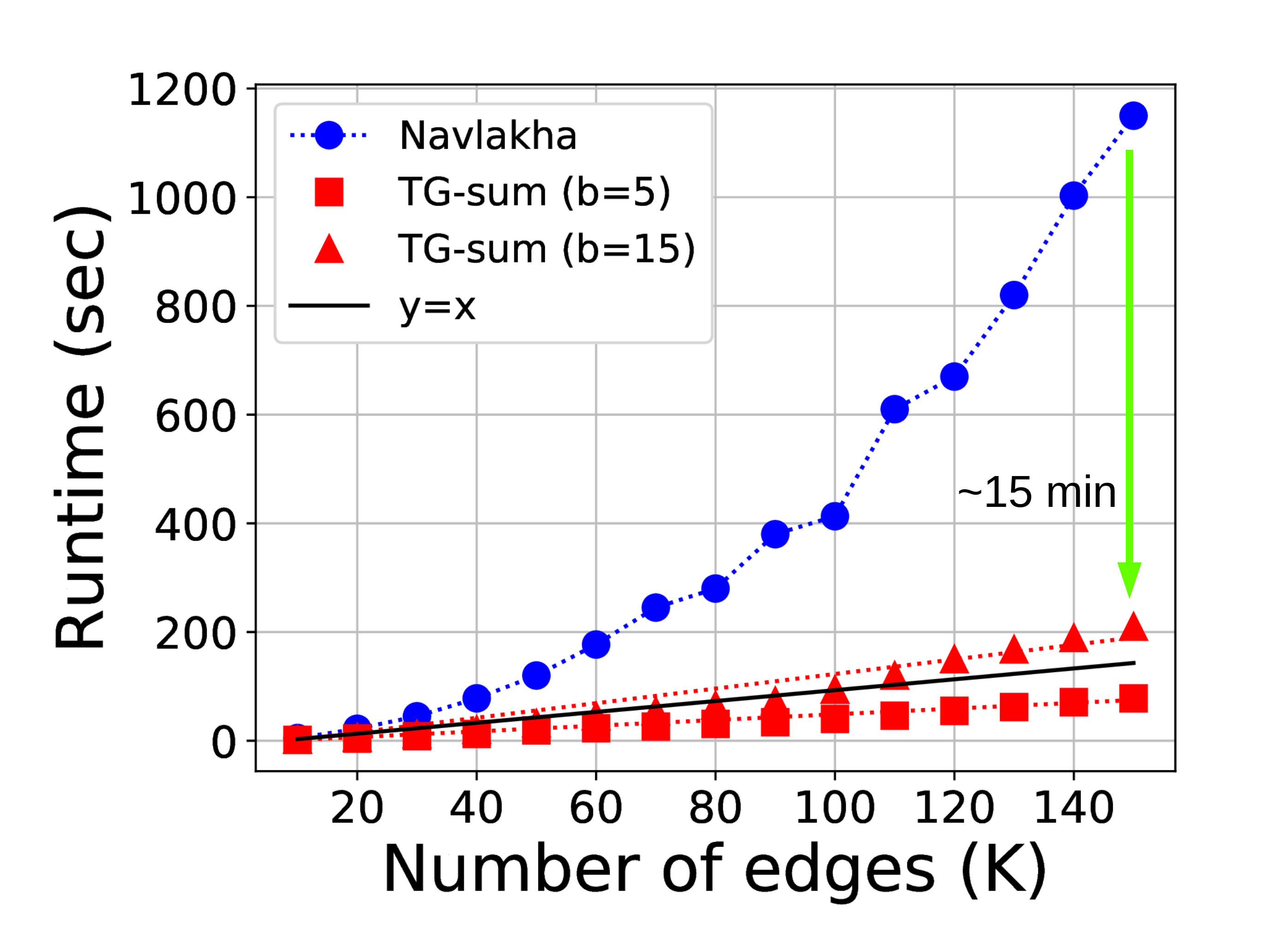}~
	\hspace{-0.3cm}\includegraphics[width=44mm]{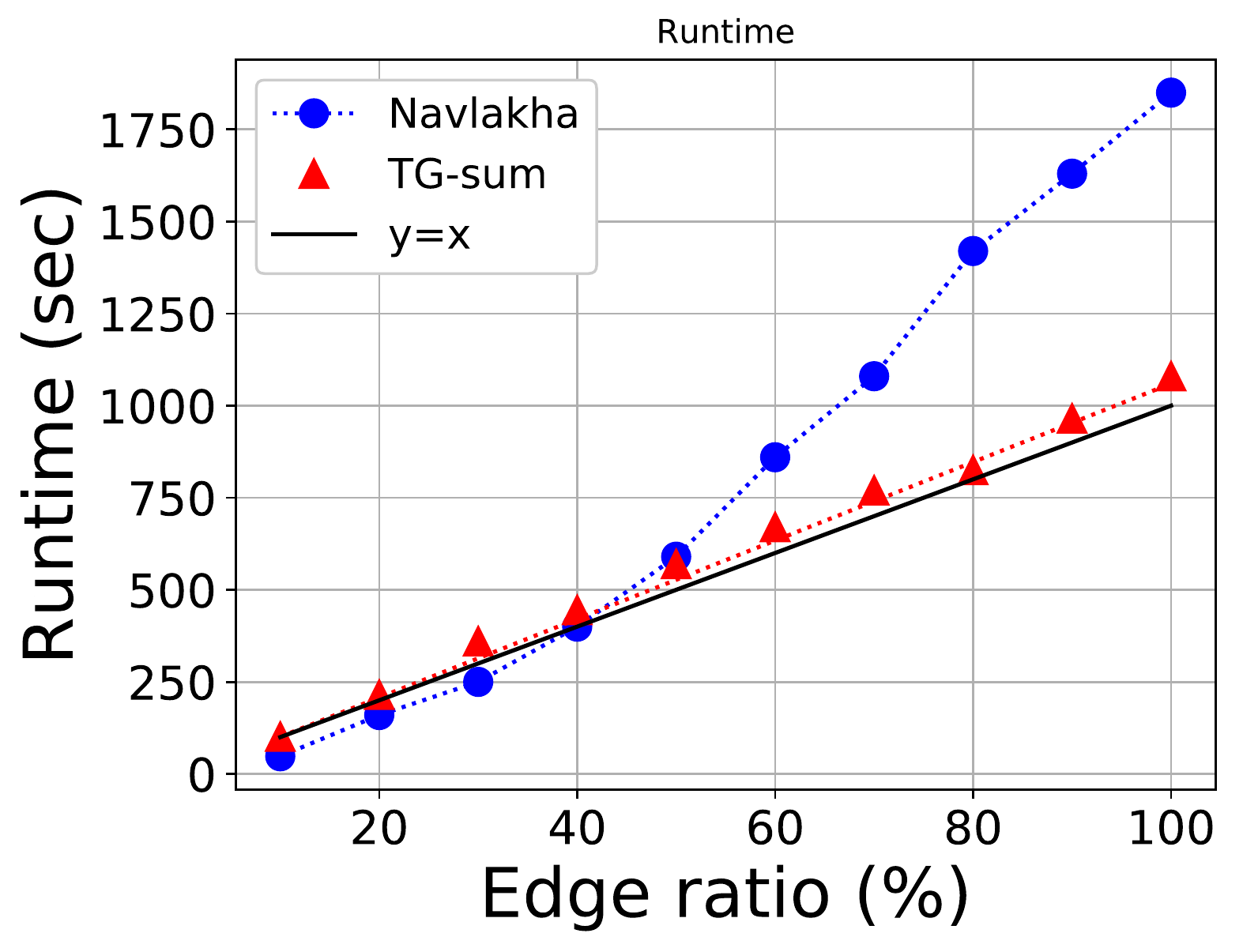}
	\vspace{-0.1in}
	%\end{tabular}
	\caption{\method scales linearly; (left) simulated graphs, (right) 4-area with increasingly sampled edges. }
	\label{fig:scale}
	\vspace{-0.3in}
\end{wrapfigure}
Results in Fig. \ref{fig:scale} (left) show that, unlike the modified Navlakha, the runtime of \method grows in a near-linear fashion. 
In addition, we sample increasing percentage of the edges uniformly at random from the 4-area graph %(our largest multi-type graph) 
to quantify scalability as a function of  number of edges where the node set is fixed. As Fig. \ref{fig:scale} (right) shows, \method is scalable to large graphs, with runtime linear in the number of edges (w/ slope $\approx 1$), %closely following the $y=x$ line with slope $1$, 
empirically confirming the complexity analysis in Sec. \ref{ssec:complexity}.

\section{Conclusion}
\label{06conclusion}

We introduced \method, a {versatile} graph summarization algorithm that ({for the first time}) can handle \textit{directed}, \textit{node-labeled}, \textit{multi}-graphs with possible self-loops (or \textit{any} combination). Built on a novel encoding scheme, % and relying on the two-part Minimum Description Length principle, 
\method seeks to minimize the total encoding cost of ($i$) a summary graph, and ($ii$) the corrections to %{losslessly} 
reconstruct the input graph losslessly.   
It efficiently finds structurally-similar nodes to create super-nodes of larger sizes incrementally, producing multi-resolution summaries. 
Extensive experiments show that \method (1) provides insights into the high-level structure of  real-world graphs, 
(2) achieves better trade-off between compression and runtime relative to baselines (only) on comparable settings, and (3) scales linearly in the number of edges.

%\input{7_acknowledgment}

\bibliographystyle{splncs04}
\bibliography{graph.bib}

\clearpage
\section{Appendix}\label{sec:sup}

\subsection{Proof of Theorem 1}
\label{ssec:thm1}
First we expand 
{{
		\begin{align*}
		L_\mathrm{ENTR} &= n(-p\log_2p -(1-p)\log_2(1-p))\\
		&=n^\prime\log_2(n/n^\prime) + (n-n^\prime)\log_2(n/(n-n^\prime))\\
		&= \log_2\frac{n^{n^\prime}}{ {n^\prime}^{n^\prime}} + \log_2\frac{n^{(n-n^\prime)}}{(n-n^\prime)^{(n-n^\prime)}}\\
		&= \log_2\frac{n^n}{ {n^\prime}^{n^\prime}(n-n^\prime)^{(n-n^\prime)}}.   
		\end{align*}
}}
Thus, to prove  the inequality in Theorem \ref{prop1},
%\eqref{eq2}, 
it suffices to show that 
{{
		\begin{align*}
		{ n \choose n^\prime } \leq \frac{n^n}{ {n^\prime}^{n^\prime}(n-n^\prime)^{(n-n^\prime)}} \;\; .
		\end{align*}	
}}
Using  Stirling's upper and lower bound on the factorial 
{{
		$$\sqrt{2\pi} n^{n+1/2}e^{-n}\geq n!\geq en^{n+1/2}e^{-n} \;, $$ 
}}
we can bound
{{
		\begin{align*}
		{ n \choose n^\prime } &= \frac{n!}{n^\prime!(n-n^\prime)!} \\
		&\leq \frac{\sqrt{2\pi}  n^n \sqrt{n} e^{-n}}{e {n^\prime}^{n^\prime} \sqrt{n^\prime} e^{-n^\prime} 
			e {(n-n^\prime)}^{n-n^\prime} \sqrt{(n-n^\prime)} e^{-(n-n^\prime)}
		}\\
		& = \frac{n^n}{ {n^\prime}^{n^\prime}(n-n^\prime)^{(n-n^\prime)}}\cdot \frac{\sqrt{2\pi} \sqrt{n}}{e^2\sqrt{n^\prime(n-n^\prime)}}\\
		&\leq \frac{n^n}{ {n^\prime}^{n^\prime}(n-n^\prime)^{(n-n^\prime)}}
		\end{align*}
}}
since 
{{
		\begin{align*}
		\frac{\sqrt{2\pi} \sqrt{n}}{e^2\sqrt{n^\prime(n-n^\prime)}}
		< \frac{ \sqrt{n}}{e\sqrt{n^\prime(n-n^\prime)}} 
		\leq \frac{ \sqrt{n}}{e\sqrt{n-1}}\leq 1 \;,
		\end{align*}
}}
for all $n > n^\prime>0$, which concludes the proof. \qed

\end{document}